\documentclass[reprint, superscriptaddress, amsmath,amssymb, aps,prd]{revtex4-2}

\RequirePackage[colorlinks=true
,urlcolor=magenta
,anchorcolor=blue
,citecolor=magenta
,filecolor=blue
,linkcolor=blue
,menucolor=blue
,linktocpage=true
,pdfproducer=medialab
,pdfa=true
]{hyperref}

\usepackage{amsmath,amssymb,amsthm,amsfonts,mathrsfs}
\usepackage{cprotect}
\usepackage{graphicx,tabularx}
\usepackage{subcaption}
\usepackage{rotating}
\usepackage{float}
\usepackage{multirow}  
\usepackage{cancel}
\usepackage{comment}
\usepackage{physics}
\usepackage{array}
\usepackage{enumitem}

\newcolumntype{P}[1]{>{\centering\arraybackslash}p{#1}}

\usepackage{xcolor}
\usepackage{dcolumn}
\usepackage{bm}

\begin{document}
\title{The Impact of Inhomogeneous Perturbations of the Inflaton on the Cosmological Primordial Magnetic Field} 

\author{Yu Li}
\email[Corresponding author: ]{leeyu@dlmu.edu.cn}
\affiliation{School of Science, Dalian Maritime University, Dalian 116026, China}

\author{Shuang Liu}
\email{1120240893shuangliu@dlmu.edu.cn}
\affiliation{School of Science, Dalian Maritime University, Dalian 116026, China}

\author{Hang Wang}
\email{wang-hang@dlmu.edu.cn}
\affiliation{School of Science, Dalian Maritime University, Dalian 116026, China}

\author{Yao-chuan Wang}
\email{ycwang@dlmu.edu.cn}
\affiliation{School of Science, Dalian Maritime University, Dalian 116026, China}

\begin{abstract}
We investigate the impact of inhomogeneous inflaton perturbations on primordial magnetic fields within the framework of generalized inflationary magnetogenesis models. Extending the Ratra model to general spacetime backgrounds, we analyze the constraint structure of the electromagnetic field and demonstrate that the standard Coulomb gauge must be generalized to accommodate spatial inhomogeneities. Instead of the vector potential, we solve the conjugate momentum with the modified initial conditions introduced by the coupling function, which become dominant during the late stages of inflation. These change the conditions under which scale-invariant electromagnetic spectra are achieved. Furthermore, we address the challenge of evaluating convolutions between vector potentials and inflaton perturbations by employing separate large- and small-scale approximations. The resulting influence to the electric and magnetic power spectra are quantified using $\Delta_E$ and $\Delta_B$, revealing a scale-dependent influence of inhomogeneities. We also find that the spectrum index evolution is sensitive to the sign of $V_{\phi}$, with distinctive behaviors for electric and magnetic fields under different scale-invariance conditions. Notably, for nearly scale-invariant magnetic fields, the perturbative effects shift the spectral index towards the red and migrate toward smaller scales as inflation progresses. We have also compared the obtained results with observational data, this offer a potential observational probe to differentiate between large-field and small-field inflation scenarios.
\begin{description}
	\item[PACS numbers]{98.80.Cq.}
\end{description}

\end{abstract}

\maketitle

\section{Introduction}
Observational evidence suggests the existence of weak large-scale magnetic fields in the universe. The typical strength of these magnetic fields is in the range of $10^{-15}{\rm G}<B_0<10^{-9}{\rm G}$ \cite{1997_barrow,1999_blasi,2009_seshadri,2010_akahori,2010_dolag,2010_neronov,2015_Planck,2019_vernstrom,2020_osullivan,2020_tiede}.
However, up to now, the origin of these magnetic fields remains an unknown and widely debated issue. Various magnetogenesis mechanisms can be broadly categorized into two types: astrophysical scenario and primordial scenario. In astrophysical scenarion, these magnetic fields originate from some astrophysical processes, which is difficult to explain the origin of the magnetic fields in cosmic voids \cite{2002_widrow,2005_hanayama,2018_safarzadeh,2021_araya,2023_papanikolaou,2023_papanikolaoua}. The magnetic fields in the cosmic voids are more likely to have originated in the very early universe \cite{2010_neronov,2017_archambault,2018_ackermann}. 

On the other hand, in primordial scenario, these large-scale magnetic fields are hypothesized to have originated in the early stages of the universe i.e. the primordial magnetic field (PMF). 
One class of possible sources of PMF are phase transitions like electroweak or QCD phase transtition in the early Universe\cite{1989_quashnock,1991_vachaspati,1993_enqvist,1994_cheng,1996_baym,1997_sigl,1998_hindmarsh,2003_kisslinger,2012_tevzadze,2019_ellis,2021_vachaspatia,2024_olea-romacho}. However, in these scenarios, only extremely weak fields form on galactic scales unless helicity is also generated—in which case an inverse cascade of energy to larger scales can occur\cite{1996_brandenburg,2004_banerjee}.

Another possible source of PMF is inflationary magnetogenesis\cite{1988_turner,1992_ratra,2008_martin,2014_kobayashi,2014_atmjeet,2015_giovannini,2015_tasinato,2016_domenech,2017_sharmaa,2018_sharma,2019_fujita,2019_shtanov,2021_bamba,2021_giovannini,2022_durrer,2022_li,2023_tripathy,2023_velasquez,2023_li,2024_dimopoulos}. Inflation provides an ideal environment for generating primordial large‑scale magnetic fields, however, the electromagnetic field cannot be amplified by inflation due to the conformal invariance of the standard electromagnetic action and the FRW metric is conformal flat. Therefore, the breaking of conformal invariance is a necessary condition for the inflationary magnetogenesis mechanism (see \cite{2010_subramanian,2011_kandus,2016_subramanian} for reviews).
A common approach to breaking conformal invariance is the Ratra model\cite{1992_ratra}, which introduces a non-minimal coupling of the form $f^2(\phi)F_{\mu\nu}F^{\mu\nu}$ between the inflaton $\phi$ and the electromagnetic field $F_{\mu\nu}$ in the Lagrangian density, where $f(\phi)$ is a time-dependent coupling function. In this paper we focus on inflationary magnetogenesis under the Ratra model. 

In the inflationary magnetogenesis mechanism, the PMF are generated during the inflationary epoch, which inevitably leads to the cosmological inhomogeneous perturbations—such as curvature perturbations and primordial gravitational wave—being affected by these PMF \cite{2007_giovannini,2012_barnaby,2013_giovanninia,2014_piccinelli,2023_adia,2025_bhaumik}. This is an effective way of linking theoretical models to observations. Conversely, cosmological inhomogeneous perturbations can also exert an influence on the PMF in the universe. This influence can be investigated by extending the Ratra model to a spacetime background with inhomogeneous perturbations. 

This issue was discussed in \cite{2001_maroto}, but in this work the standard Coulomb gauge has been adopted. However, in a spacetime background with spatial inhomogeneities, the standard Coulomb gauge does not automatically satisfy the Gauss constraint for the electromagnetic field as it does in the FRW metric (see \cite{1992_tolksdorf,2004_tsagas,2023_hwang,2024_egorov} for example). The impact of inhomogeneous perturbations on the Coulomb gauge, in turn, affects inflationary magnetogenesis models. Therefore, this effect should be taken into account. There are two possible approaches to address this issue. One is to use the standard Coulomb gauge in combination with the Gauss constraint to restrict the form of the coupling function. The other is to adopt a generalized Coulomb gauge, thereby allowing more freedom in the choice of the coupling function. The former approach imposes constraints on the coupling function that are unfavorable for constructing viable inflationary magnetogenesis models \cite{2022_li}. Therefore, in this paper, we adopt the latter approach.

This paper is organized as follow: In Section \ref{s2}, we extend the Ratra model to an arbitrary spacetime background and present the Maxwell equations within the Lagrangian framework. We then discuss the constrained system of the electromagnetic field in the Hamiltonian framework; In Section \ref{s3}, we construct the inflationary magnetogenesis model in the presence of both metric perturbations and inflaton field perturbations. We then perform the quantization of the electromagnetic field and finally solve for the mode functions; The power spectrum of the electromagnetic field will be derived in Section \ref{s4}, along with the impact of inhomogeneous perturbations on the spectrum; In Section \ref{sa1}, we will discuss the impact of inhomogeneous perturbations on the spectral index of the primordial magnetic field in different inflationary models and compare the results with the Planck 2018 data; Summary  is presented in Section \ref{s5}.
\section{Ratra model in general spacetime \label{s2}}
\subsection{Lagrangian formalism}
Firstly, consider the general space-time region $M$ with metric $g_{\mu\nu}$.
The action is 
\begin{equation}
	\label{e1}
	S=\int\sqrt{-g}d^4x\left(\mathcal{L}_g+\mathcal{L}_{\phi}+\mathcal{L}_{\rm EM}\right)
\end{equation}
where 
\begin{align}
	\mathcal{L}_g:=&\frac{1}{16\pi G}R\hskip 5pt(\text{gravitational field})\label{e2}\\
	\mathcal{L}_{\phi}:=&-\frac{1}{2}g^{\mu\nu}(\nabla_{\mu}\phi)(\nabla_{\nu}\phi)-V(\phi)\hskip 5pt(\text{scalar field})\label{e3}\\
	\mathcal{L}_{\rm EM}:=&-\frac{f^2(\phi)}{16\pi}F_{\mu\nu}F^{\mu\nu}\hskip 5pt(\text{electromagnetic field})\label{e4}
\end{align}
$f(\phi)$ is coupling function between the scalar field $\phi$ and the electromagnetic field $F_{\mu\nu}:=\partial_{\mu}A_{\nu}-\partial_{\nu}A_{\mu}$, where$A_{\mu}$ is the dual 4-vector of electromagnetic field.
By considering the variation of the action \eqref{e1} with regard to $A_{\mu}$ and utilizing Hamilton's principle, the evolution equations of the electromagnetic field can be derived as follows:
\begin{equation}
	\label{e5}
	\nabla^{\mu}\left[f^2(\phi)F_{\mu\nu}\right]=0
\end{equation}
On the other hand, noticed that $F_{\mu\nu}$ is closed 2-form, then
\begin{equation}
	\label{e6}
	\nabla^{\mu}\tilde{F}_{\mu\nu}=0
\end{equation}
where $\tilde{F}_{\mu\nu}:=\frac{1}{2}F^{\rho\lambda}\epsilon_{\mu\nu\rho\lambda}$ is dual form of $F_{\mu\nu}$ 
and $\epsilon_{\mu\nu\rho\lambda}$ is volume element which is compatible with the metric of spacetime.
\eqref{e5} and \eqref{e6} together form Maxwell's equations.

The evolution equation of scalar field (Klein-Gordon equation) can be obtained by take variation of action \eqref{e1} with $\phi$:
\begin{equation}
	\label{e7}
	\frac{1}{\sqrt{-g}}\partial_{\nu}\left(\sqrt{-g}g^{\mu\nu}\partial_{\mu}\phi\right)-V_{\phi}=
	\frac{ff_{\phi}}{8\pi}F_{\mu\nu}F^{\mu\nu}
\end{equation}

where $f_{\phi}:=df/d\phi,~V_{\phi}:=dV/d\phi$.

From \eqref{e7}, it can be seen that the electromagnetic field can serve as a source for the scalar field. However, \eqref{e5} indicates that the scalar field is not a source for the electromagnetic field. Nevertheless, \eqref{e5} also shows that the evolution of the scalar field still influences the evolution of the electromagnetic field through the coupling function $f(\phi)$.
\subsection{Hamiltonian formalism}
To better analyze the constraint structure of the system, we will recast the content of the previous subsection into the Hamiltonian form in this subsection.

To get the Hamiltanian, the spacetime should be 3+1 decomposed. Choose a timelike vector field $Z^{\mu}$ as the 
observer 4-velocity, and the 3+1 decomposition is 
\begin{equation}
	\label{e8}
	Z^{\mu}=Nn^{\mu}+N^{\mu}
\end{equation}
where $N$ is lapse function, $N^{\mu}$ is the shift vector field and $n^{\mu}$ is space hypersurface orthogonal timelike unit vector.
The metirc of space hypersurface is definded by $h_{\mu\nu}:=g_{\mu\nu}+n_{\mu}n_{\nu}$. In this paper, we choose configuration variables as:
\begin{equation}
	\label{e9}
	\mathbb{A}_0:=Z^{\mu}A_{\mu},\hskip 10pt \mathbb{A}_{\mu}:=h^{\nu}_{\mu}A_{\nu}
\end{equation}
where $h^{\nu}_{\mu}=\delta^{\nu}_{\mu}+n^{\nu}n_{\mu}$ is spatial projection operator.Using the above configuration variables, the Lagrangian density of the electromagnetic field can be expressed in the form of a 3+1 decomposition:
\begin{widetext}
	\begin{align}
		\sqrt{-g}\mathcal{L}_{\rm EM}=&-\frac{N\sqrt{h}f^2(\phi)}{16\pi}\left[-\frac{2}{N^2}h^{\mu\nu}(D_{\mu}\mathbb{A}_0-D_0\mathbb{A}_{\mu})(D_{\nu}\mathbb{A}_0-D_0\mathbb{A}_{\nu})+\frac{4}{N^2}h^{\mu\nu}N^{\rho}(D_{\mu}\mathbb{A}_0-D_0\mathbb{A}_{\mu})(D_{\nu}\mathbb{A}_{\rho}-D_{\rho}\mathbb{A}_{\nu})\right.\nonumber\\
		&\left.+\left(h^{\mu\nu}h^{\rho\sigma}-\frac{2}{N^2}h^{\mu\nu}N^{\rho}N^{\sigma}\right)(D_{\mu}\mathbb{A}_{\nu}
		-D_{\nu}\mathbb{A}_{\mu})(D_{\rho}\mathbb{A}_{\sigma}-D_{\sigma}\mathbb{A}_{\rho})\right]
		\label{e10}
	\end{align}
\end{widetext}
where $D_{\mu}$ is the spatial derivative operator which is compatible with $h_{\mu\nu}$, and $D_0$ is time derivative operator which is 
definded by
\begin{equation}
	\label{e11}
	D_0\mathbb{A}_{\mu}:=h^{\nu}_{\mu}\mathscr{L}_{Z}A_{\nu}
\end{equation}
$\mathscr{L}_Z$ is Lie derivative along $Z^{\mu}$. One can see that, the Lagrangian density \eqref{e10} is the local functional of 
$\mathbb{A}_0,\mathbb{A}_{\mu}$ and their spatial and time derivatives. However, it independ on $D_0\mathbb{A}_0$, which means that, 
there is a primary constraint. The canonical momentum can be defined as:
\begin{align}
		\Pi_A^0:=&\frac{\partial (\sqrt{-g}\mathcal{L}_{\rm EM})}{\partial (D_0\mathbb{A}_0)}=0\label{e12}\\
		\Pi_A^{\mu}:=&\frac{\partial (\sqrt{-g}\mathcal{L}_{\rm EM})}{\partial (D_0\mathbb{A}_{\mu})}\nonumber\\
				=&-\frac{\sqrt{h}f^2(\phi)}{4\pi N}(D^{\mu}\mathbb{A}_0-h^{\mu\nu}D_0\mathbb{A}_{\nu}-h^{\mu\nu}N^{\rho}\mathbb{F}_{\nu\rho})
				\label{e13}
\end{align}
where
\begin{equation}
	\label{e14}
	\mathbb{F_{\nu\rho}}:=h^{\mu}_{\nu}h^{\sigma}_{\rho}F_{\mu\sigma}=D_{\nu}\mathbb{A}_{\rho}-D_{\rho}\mathbb{A}_{\nu}
\end{equation}
\eqref{e12} is the primary constraint. Through the Legendre transformation, the Hamiltonian of electromagnetic field can be obtained as:
\begin{equation}
	\label{e15}
	H_{\rm EM}=\int \sqrt{h}d^3x\left(N\mathscr{H}_0+N^{\mu}\mathscr{H}_{\mu}+\mathbb{A}_0\mathscr{G}\right)
\end{equation}
where
\begin{align}
	\mathscr{H}:=&\frac{2\pi}{hf^2}h_{\mu\nu}\Pi_A^{\mu}\Pi_A^{\nu}+\frac{f^2}{16\pi}h^{\mu\rho}h^{\nu\sigma}\mathbb{F}_{\mu\nu}\mathbb{F}_{\rho\sigma}\label{e16}\\
	\mathscr{H}_{\mu}:=&\frac{1}{\sqrt{h}}\Pi_A^{\nu}\mathbb{F}_{\mu\nu}\label{e17}\\
	\mathscr{G}:=&-D_{\mu}\left(\frac{1}{\sqrt{h}}\Pi_A^{\mu}\right)\label{e18}
\end{align}
The Hamiltonian canonical equations yield Maxwell's equations:
\begin{subequations}
	\begin{align}
		D_0\mathbb{A}_0=&\lambda\label{e19a}\\
		D_0\Pi_A^0=&-\sqrt{h}\mathscr{G}\label{e19b}\\
		D_0\mathbb{A}_{\mu}=&\frac{4\pi N}{\sqrt{h}f^2}h_{\mu\nu}\Pi_A^{\nu}-N^{\nu}\mathbb{F}_{\mu\nu}+D_{\mu}\mathbb{A}_0\label{e19c}\\
		D_0\Pi_A^{\mu}=&\frac{\sqrt{h}}{4\pi}D_{\rho}(Nf^2h^{\mu\nu}h^{\rho\sigma}\mathbb{F}_{\nu\sigma})
		+2\sqrt{h}D_{\nu}\left(\frac{1}{\sqrt{h}}N^{[\nu}\Pi_A^{\mu]}\right)\label{e19d}
	\end{align}
\end{subequations}
where $\lambda$ is Lagrange multiplier. Primary constraint \eqref{e12} must satisfy the consistency condition $D_0\Pi_A^0=0$, 
using \eqref{e19b} one can get the sencondary constraint, which is known as Gauss constraint:
\begin{equation}
	\label{e20}
\mathscr{G}=0~\Rightarrow~D_{\mu}\left(\frac{\Pi_A^{\mu}}{\sqrt{h}}\right)=0
\end{equation}

\subsection{Constraint algebra}
In order to analyze the constraint structure of the system, it is necessary to define constraint functionals for primary and secondary constraints separately. Let $\xi,\chi$ be an arbitrary time-independent scalar fields on a spacelike hypersurface $\Sigma_t$ that satisfies appropriate conditions, then the primary and secondary constraint functianls are defined by
\begin{subequations}
	\begin{align}
		\mathcal{C}_{\xi}:=&\int_{\Sigma_t}\Pi_A^0\xi\label{e21a}\\
		\mathcal{C}_{\chi}:=&-\int_{\Sigma_t}\chi\mathscr{G}=\int_{\Sigma_t}\left[\chi D_{\mu}\left(\frac{\Pi_A^{\mu}}{\sqrt{h}}\right)\right]\label{e21b}
	\end{align}
\end{subequations}
One can check that $D_0\mathcal{C}_{\chi}=0$, which means that there is no other constraint. It can also be verified that the Poisson brackets between the constraint functionals are:
\begin{equation}
	\label{e22}
	\left\{\mathcal{C}_{\chi},\mathcal{C}_{\chi'}\right\}=\left\{\mathcal{C}_{\xi},\mathcal{C}_{\xi'}\right\}
	=\left\{\mathcal{C}_{\chi},\mathcal{C}_{\xi}\right\}=0
\end{equation}
This implies that all the constraints are of the first class, which means that $\lambda$ is free Lagrange multiplier and can be 
set to zero. From \eqref{e19a}, this setting gives $D_0\mathbb{A}_0=0$, i.e. $\mathbb{A}_0=\text{constant}$.
By appropriately choosing the value of $\mathbb{A}_0$ and noting that $\Pi_A^0=0$, the pair of canonical variables $(\mathbb{A}_0,\Pi_A^0)$ is no longer present in the dynamical system, and thus they can be removed from the set of canonical variables.

Let the remaining phase space be $\bar{\Gamma}$, with a dimension of $6\times\infty^3$. Only secondary constraints remain on it, and the constraint vector field corresponding to these constraints can be expanded in the canonical coordinates $(\mathbb{A}_{\mu},\Pi_A^{\mu})$ basis as:
\begin{align}
	X_{\mathcal{C}_{\chi}}^{\alpha}=&\int _{\Sigma_t}\left[\frac{\delta\mathcal{C}_{\chi}}{\delta\Pi_A^{\mu}}\left(\frac{\delta}{\delta\mathbb{A}_{\mu}}\right)^{\alpha}
	-\frac{\delta\mathcal{C}_{\chi}}{\delta\mathbb{A}_{\mu}}\left(\frac{\delta}{\delta\Pi_A^{\mu}}\right)^{\alpha}\right]\nonumber\\
	=&-\int_{\Sigma_t}D_{\mu}\chi\left(\frac{\delta}{\delta\mathbb{A}_{\mu}}\right)^{\alpha}
	\label{e23}
\end{align}
where $\alpha$ is index of vector on $\bar{\Gamma}$. The infinitesimal gauge transformation of the electromagnetic field is given by the diffeomorphism along the integral curves of the vector field $X_{\mathcal{C}_{\chi}}^{\alpha}$ as:
\begin{equation}
	\label{e24}
	(\mathbb{A}_{\mu},\Pi_A^{\mu}) \mapsto (\mathbb{A}_{\mu}-\epsilon D_{\mu}\chi,\Pi_A^{\mu})
\end{equation}
It can be verified that the electric and magnetic fields remain invariant under the above gauge transformation.

The constraint $\mathcal{C}_{\chi}$ reduces the $\bar{\Gamma}$ to $\bar{\Gamma}'$ which dimension is $5\times \infty^3$.
Constraint vector field $X_{\mathcal{C}_{\chi}}^{\alpha}$ tangent to $\bar{\Gamma}'$, which mean that the intergral curves of $X_{\mathcal{C}_{\chi}}^{\alpha}$ lie on $\bar{\Gamma}'$. Each point on an integral curve represents the same physical state, so these points can be defined as an equivalence class. The set of these equibalence classes is the reduced phase spce $\tilde{\Gamma}$ with a dimension of $4\times \infty^3$. This corresponds to the two independent components of the electromagnetic vector field.

The reduced phase space $\tilde{\Gamma}$ can also be constructed by selecting an appropriate scalar field $\chi$ (selecting a gauge), thereby imposing certain conditions on $\mathbb{A}_{\mu}-\epsilon D_{\mu}\chi$. 

A commonly used gauge is to choose $\chi$ such that $D^{\mu}(\mathbb{A}_{\mu}-\epsilon D_{\mu}\chi)=0$, meaning $\chi$ must satisfy $\epsilon D^{\mu}D_{\mu}\chi=D^{\mu}\mathbb{A}_{\mu}$. This is known as the Coulomb gauge. Under this selection, the constraint $\mathcal{C}_{\chi}$ is
\begin{align}
	\label{e25}
	\mathcal{C}_{\chi}=&\int_{\Sigma_t}\chi\left\{
	\frac{f^2}{4N\pi}\left[N^{\nu}D^{\mu}D_{\nu}\mathbb{A}_{\mu}-N^{\nu}D^{\mu}D_{\mu}\mathbb{A}_{\nu}\right.\right.\nonumber\\
	&\left.+(D_{\nu}\mathbb{A}_{\mu})(D^{\mu}N^{\nu})-(D_{\mu}\mathbb{A}_{\nu})(D^{\mu}N^{\nu})\right]\nonumber\\
	&\left.+D^{\mu}\left(\frac{f^2}{4N\pi}\right)(D_0\mathbb{A}_{\mu}-N^{\nu}\mathbb{F}_{\mu\nu})
	\right\}
\end{align}
Obviously, the above constraint vanishes automatically only when the spatial hypersurface is homogeneous. However, for a general homogeneous spatial case, such as when considering spatial inhomogeneities, the standard Coulomb gauge does not automatically satisfy the constraint. This effectively treats physical states that are no longer on the same integral curve as an equivalence class.

There are two ways to solve this problem:  
\begin{enumerate}
	\item Choosing an appropriate coupling function $f(\phi)$ so that the constraint is automatically satisfied in the Coulomb gauge. However, this restricts the form of the coupling function—it must ensure the consistency of the theory.  
	\item  Selecting a new gauge that automatically satisfies the constraint. This approach is equivalent to solving the constraint equation together with Maxwell’s equations.  
\end{enumerate}
In this paper, we adopt the second approach.
\section{Inflationary magnetogenesis \label{s3}}
In this section, we consider the inflationary magnetogenesis under the FRW metric with inhomogeneous perturbations:
\begin{equation}
	\label{e26}
	g_{\mu\nu}=\bar{g}_{\mu\nu}+\gamma_{\mu\nu}
\end{equation}
where $\bar{g}_{\mu\nu}$ is homogeneous and isotropic FRW metric, and its components in the standard 3+1 decomposition are:
\begin{equation}
	\label{e27}
	\bar{g}_{00}=-1,~\bar{g}_{i0}=\bar{g}_{0i}=0,~ \bar{g}_{ij}=a^2(t)\delta_{ij}
\end{equation}
where $a$ is scale factor and $t$ is cosmic time. $\gamma_{\mu\nu}$ represents the scalar mode inhomogeneous metric perturbation. Its components in the conformal Newtonian gauge are:
\begin{equation}
	\label{e28}
	\gamma_{00}=-2\Psi({\bm x},t),~ \gamma_{0i}=\gamma_{i0}=0,~\gamma_{ij}=2a^2\Phi({\bm x},t)\delta_{ij}
\end{equation}
The inflaton can also be decomposed into a homogeneous part and an inhomogeneous perturbation part:
\begin{equation}
	\label{e29}
	\phi({\bm x},t)=\bar{\phi}(t)+\delta\phi({\bm x},t)
\end{equation}
Thus, the coupling function $f(\phi)$ can be expanded according to the inhomogeneous perturbation as:
\begin{equation}
	\label{e30}
	f(\phi)=\bar{f}+\theta\delta\phi+\cdots
\end{equation}
where
\begin{equation}
	\label{e31}
	\bar{f}:=f(\bar{\phi}),\hskip 10pt \theta:=\frac{df}{d\phi}\Big|_{\phi=\bar{\phi}}
\end{equation}
It is worth noting that $\bar{f}$ and $\theta$ only depend on time.
\subsection{Evolution equation}
Using the metric \eqref{e26}  one can get the time component of \eqref{e5}:
\begin{equation}
	\label{e32}
	a\bar{f}^2\delta^{ij}\partial_i\left[\left(1+\Phi-\Psi+\frac{2\theta}{\bar{f}}\delta\phi\right)D_0\mathbb{A}_j\right]=0
\end{equation}
It is easy to verify that \eqref{e32} is exactly the constraint \eqref{e20}. If $\Phi=\Psi=\delta\phi=0$, \eqref{e32} 
is the standard Coulomb gauge. However, after considering inhomogeneous perturbations, \eqref{e32} does not give the standard Coulomb gauge, which is consistent with the discussion in the previous section. If conformal time $\eta:=\int a^{-1} dt$ is chosen, and considering $\Phi=-\Psi$, then the constraint equation becomes:
\begin{equation}
	\label{e33}
	\delta^{ij}\partial_i\left[\left(1-2\Psi+\frac{2\theta}{\bar{f}}\delta\phi\right)\mathbb{A}_j'\right]=0
\end{equation}
where ``$\prime$" denotes the derivative with respect to conformal time. The spatial componets of \eqref{e5} are:
\begin{align}
	\label{e34}
	\mathbb{A}_i''&+2\left[\frac{\bar{f}'}{\bar{f}}-\Psi'+\left(\frac{\theta}{\bar{f}}\delta\phi\right)'\right]\mathbb{A}_i'\nonumber\\
	&-2\delta_{\ell k}\left(\Psi_{,\ell}+\frac{\theta}{\bar{f}}\delta\phi_{,\ell}\right)(\mathbb{A}_{i,k}-\mathbb{A}_{k,i})\nonumber\\
	&-(1+4\Psi)\delta^{\ell k}(\partial_{\ell}\partial_k\mathbb{A}_i-\partial_{\ell}\partial_i\mathbb{A}_k)=0
\end{align}
It is not easy to solve equations \eqref{e33} and \eqref{e34} simultaneously, however one can notice that, constraint equation
can be rewirten as:
\begin{equation}
	\label{e35}
	\partial_i\bar{\Pi}_A^i=0
\end{equation}
where 
\begin{equation}
	\label{e36}
	\bar{\Pi}_A^i:=\frac{4\pi}{\bar{f}^2}\Pi_A^i=\left(1-2\Psi+\frac{2\theta}{\bar{f}}\delta\phi\right)\delta^{ij}\mathbb{A}_j'
\end{equation}
Equation \eqref{e35}  is similar to the standard Coulomb gauge condition satisfied by $\mathbb{A}_i$. Therefore, we can first solve for $\bar{\Pi}_A^i$, and then use equation \eqref{e36} to solve for $\mathbb{A}_i$. Using \eqref{e34}, one can derive the equation that $\bar{\Pi}_A^i$ satisfies as:
\begin{widetext}
\begin{equation}
		\label{e37}
		\delta_{ij}(\bar{\Pi}_A^j)''+2\frac{\bar{f}'}{\bar{f}}\delta_{ij}(\bar{\Pi}_A^j)'
		+2\left[\frac{\bar{f}''}{\bar{f}}-\left(\frac{\bar{f}'}{\bar{f}}\right)^2\right]\delta_{ij}\bar{\Pi}_A^j-\delta^{\ell m}\delta_{ij}\bar{\Pi}_{A,\ell m}^j
		=\mathcal{Q}_i
\end{equation}
where
\begin{align}
	\mathcal{Q}_i=&4\frac{\theta}{\bar{f}}\delta\phi\delta^{\ell m}\delta_{ij}\bar{\Pi}_{A,\ell m}^j+\left(-2\Psi_{,\ell}+6\frac{\theta}{\bar{f}}\delta\phi_{,\ell}\right)\delta^{\ell m}\delta_{ij}\bar{\Pi}_{A,m}^j-4\frac{\theta}{\bar{f}}\delta\phi_{,j}\bar{\Pi}_{A,i}^j\nonumber\\
	&+\left[2\Psi'+\left(\frac{2\theta}{\bar{f}}\right)'\delta\phi+\frac{2\theta}{\bar{f}}\delta\phi'\right]
	\int d\tau\delta^{\ell m}\delta_{ij}\bar{\Pi}_{A,\ell m}^j+\delta^{\ell m}\left[2\Psi_{,\ell}'+\left(\frac{2\theta}{\bar{f}}\right)
	\delta\phi_{,\ell}+\frac{2\theta}{\bar{f}}\delta\phi_{,\ell}'\right]\int d\tau \left(\delta_{ij}\bar{\Pi}_{A,m}^j-\delta_{mj}\bar{\Pi}_{A,i}^j\right)\nonumber\\
	&-\left[\delta^{\ell m}\delta_{ij}\left(2\Psi_{,\ell m}-\frac{2\theta}{\bar{f}}\delta\phi_{,\ell m}\right)-\left(2\Psi_{,ij}-\frac{2\theta}{\bar{f}}\delta\phi_{,ij}\right)\right]\bar{\Pi}_A^j\label{e38}
\end{align}
\end{widetext}
From equation \eqref{e38}, it can be seen that $\mathcal{Q}_i\sim O(\delta\phi)$, so it can be neglected compared to the terms on the left-hand side of the equation. Therefore, we will only discuss the case where $\mathcal{Q}_i\sim 0$ in the following analysis. Under this consideration, equation \eqref{e38} simplifies to:
\begin{equation}
	\label{e39}
	\bar{\Pi}_A^{i~\prime\prime}+2\frac{\bar{f}'}{\bar{f}}\bar{\Pi}_A^{i~\prime}+2\left[\frac{\bar{f}''}{\bar{f}}-\left(\frac{\bar{f}'}{\bar{f}}\right)^2\right]\bar{\Pi}_A^i-\delta^{jk}\bar{\Pi}_{A,jk}^i=0
\end{equation}

\subsection{Quantization of electromagnetic field}
Performing a plane wave expansion for $\bar{\Pi}_A^i$ and $\mathbb{A}_i$:
\begin{subequations}
	\begin{align}
		\bar{\Pi}_A^i&=\int\frac{d^3{\bm k}}{(2\pi)^3}\sum_{\lambda}\left[\varpi b_{\lambda} e^{i{\bm k}\cdot{\bm x}}
		+\varpi^*b_{\lambda}^{\dagger}e^{-i{\bm k}\cdot{\bm x}}\right]{\bm e}_{\lambda}^i\label{e40a}\\
		\mathbb{A}_i=&\int\frac{d^3{\bm k}}{(2\pi)^3}\sum_{\lambda}\left[\mathcal{A} b_{\lambda} e^{i{\bm k}\cdot{\bm x}}
		+\mathcal{A}^*b_{\lambda}^{\dagger}e^{-i{\bm k}\cdot{\bm x}}\right]h_{ij}{\bm e}_{\lambda}^j\label{e40b}
	\end{align}
\end{subequations}
where $\varpi({\bm k},\eta)$ and $\mathcal{A}({\bm k},\eta)$ are the mode function of $\bar{\Pi}_A^i$ and $\mathbb{A}_i$, respectively.
According to the \eqref{e36} and  convolution theorem of the Fourier transform, the relationship between the two mode functions can be derived as:
\begin{equation}
	\label{e41}
	\left[a\mathcal{A}({\bm k},\eta)\right]'=\int\frac{d^3{\bm \kappa}}{(2\pi)^3}
	\left[\delta({\bm \kappa})+2\Psi_{\bm \kappa}-\frac{2\theta}{\bar{f}}\delta\phi_{\bm\kappa}\right]\varpi({\bm k}-{\bm \kappa},\eta)
\end{equation}
where $\Psi_{\bm \kappa}$ and $\delta\phi_{\bm\kappa}$ are Fourier transform of $\Psi$ and $\delta\phi$, respectively.
At the beginning of inflation, inhomogeneous perturbations can be ignored, and at this time, \eqref{e41} becomes:
\begin{equation}
	\label{e52}
	[a\mathcal{A}({\bm k},\eta)]'=\varpi({\bm k},\eta)
\end{equation}

The energy densities of the electric and magnetic fields can be derived from the Hamiltonian \eqref{e15} as:
\begin{subequations}
	\begin{align}
		\hat{\rho}_E&=\frac{\bar{f}^2}{8\pi a^4}\left(1+4\Psi-2\frac{\theta}{\bar{f}}\delta\phi\right)\delta_{ij}\bar{\Pi}_A^i\bar{\Pi}_A^j\label{e42a}\\
		\hat{\rho}_B&=\frac{\bar{f}^2}{4\pi a^4}\left(1+4\Psi+\frac{2\theta}{\bar{f}}\delta\phi\right)\delta^{i[\ell}\delta^{m]j}\mathbb{A}_{j,i}\mathbb{A}_{m,\ell}\label{e42b}
	\end{align}
\end{subequations}
Their vacuum expectation values can be written as:
\begin{subequations}
	\begin{align}
		\rho_E=&\langle\hat{\rho}_E\rangle=\frac{\bar{f}^2}{8\pi a^4}\langle\delta_{ij}\hat{\Pi}_A^i\hat{\Pi}_A^j\rangle\label{e43a}\\
		\rho_B=&\langle\hat{\rho}_B\rangle=\frac{\bar{f}^2}{4\pi a^4}\langle\delta^{i[\ell}\delta^{m]j}\hat{\mathbb{A}}_{ij}\hat{\mathbb{A}}_{\ell m}\rangle\label{e43b}
	\end{align}
\end{subequations}
where
\begin{subequations}
	\begin{align}
		\hat{\Pi}_A^i:=&\left(1+2\Psi-\frac{\theta}{\bar{f}}\delta\phi\right)\bar{\Pi}_A^i\label{e44a}\\
		\hat{\mathbb{A}}_{ij}:=&\left(1+2\Psi+\frac{\theta}{\bar{f}}\delta\phi\right)\mathbb{A}_{j,i}\label{e44b}
	\end{align}
\end{subequations}
From \eqref{e40a} and \eqref{e40b}, one can get the wave expansion for $\hat{\Pi}_A^i$ and $\hat{\mathbb{A}}_i$:
\begin{subequations}
	\begin{align}
		\hat{\Pi}_A^i=&\int\frac{d^3{\bm k}}{(2\pi)^3}\sum_{\lambda}\left[\hat{\varpi}b_{\lambda}e^{i{\bm k}\cdot{\bm x}}
		+\hat{\varpi}^*b_{\lambda}^{\dagger}e^{-i{\bm k}\cdot{\bm x}}\right]{\bm e}_{\lambda}^i\label{e45a}\\
		\hat{\mathbb{A}}_{ij}=&\int\frac{d^3{\bm k}}{(2\pi)^3}\sum_{\lambda}\left[\hat{\mathcal{A}}_ib_{\lambda}e^{i{\bm k}\cdot{\bm x}}
		+\hat{\mathcal{A}}_i^*b_{\lambda}^{\dagger}e^{-i{\bm k}\cdot{\bm x}}\right]h_{jm}{\bm e}^m_{\lambda}\label{e45b}
	\end{align}
\end{subequations}
where the mode functions of $\hat{\Pi}_A^i$ and $\hat{\mathbb{A}}_{ij}$ are
\begin{subequations}
	\begin{align}
		\hat{\varpi}=&\int \frac{d^3{\bm \kappa}}{(2\pi)^3}\left[\delta({\bm \kappa})+2\Psi_{\bm \kappa}-
		\frac{\theta}{\bar{f}}\delta\phi_{\bm \kappa}\right]\varpi({\bm k}-{\bm\kappa},\eta)\label{e46a}\\
		\hat{\mathcal{A}}_i=&a\int\frac{d^3{\bm\kappa}}{(2\pi)^3}\left[\delta({\bm \kappa})+\frac{\theta}{\bar{f}}\delta\phi_{\bm\kappa}\right](k_i-\kappa_i)\mathcal{A}({\bm k}-{\bm \kappa},\eta)\label{e46b}
	\end{align}
\end{subequations}
Using \eqref{e46a} and \eqref{e46b}, \eqref{e43a} and \eqref{e43b} can be rewirtten as:
\begin{subequations}
	\begin{align}
		\rho_E=&\int d\ln k\frac{\bar{f}^2k^3}{8\pi^3a^4}\left|\hat{\varpi}\right|^2\label{e47a}\\
		\rho_B=&\int d\ln k\frac{\bar{f}^2k^3}{8\pi^3a^4}\left|\hat{\mathcal{A}}\right|^2\label{e47b}
	\end{align}
\end{subequations}
where $\left|\hat{\mathcal{A}}\right|^2:=\delta^{ij}\hat{\mathcal{A}}_i\hat{\mathcal{A}}_j^*$. Therefore the power 
spectrum of electromagnetic field are 
\begin{subequations}
	\begin{align}
		\mathscr{P}_E=&\frac{\bar{f}^2k^3}{8\pi^3a^4}\left|\hat{\varpi}\right|^2\label{e48a}\\
		\mathscr{P}_B=&\frac{\bar{f}^2k^3}{8\pi^3a^4}\left|\hat{\mathcal{A}}\right|^2\label{e48b}
	\end{align}
\end{subequations}
At this point, calculating the above power spectrum reduces to solving for mode function $\varpi({\bm k},\eta)$.

\subsection{Solution of mode function $\varpi$}
Insert \eqref{e40a} into \eqref{e39} one can get the evolution equation of mode function $\varpi$:
\begin{equation}
	\label{e49}
	\varpi''+2\frac{\bar{f}'}{\bar{f}}\varpi'+2\left[\frac{\bar{f}''}{\bar{f}}-\left(\frac{\bar{f}'}{\bar{f}}\right)^2\right]\varpi
	+k^2\varpi=0
\end{equation}
In this paper, we consider the power law coupling function $\bar{f}\propto \eta^{\gamma}$, the general solution of Equation \eqref{e49} under this consideration is:
\begin{equation}
	\label{e50}
	\varpi(k,\eta)=\frac{\sqrt{-k\eta}}{\bar{f}}\left[C_1J_{\gamma+1/2}(-k\eta)+C_2J_{-\gamma-1/2}(-k\eta)\right]
\end{equation}
where $J$ is Bessel function and $C_1,C_2$ need to be determined by the initial conditions.

In typical inflationary magnetogenesis models, the BD vacuum initial condition is usually adopted for mode function $\mathcal{A}$ \cite{2010_subramanian} :
\begin{equation}
	\label{e51}
	\mathcal{A}\to \frac{1}{a\bar{f}}\sqrt{\frac{2\pi}{k}}e^{-i k\eta}
\end{equation}
From \eqref{e52}, one can get the initial condition for $\varpi$:
\begin{equation}
	\label{e53}
	\varpi\to\frac{1}{\bar{f}}\sqrt{\frac{2\pi}{k}}\left(-\frac{\bar{f}'}{\bar{f}}-ik\right)e^{-ik\eta}
\end{equation}
It is worth noting that the first term inside the bracket in the above condition i.e. $\bar{f}'/\bar{f}$ is proportional to $\eta^{-1}$. This indicates that at the beginning of inflation, this term can be ignored. However, as inflation progresses, its influence grows increasingly significant. From the subsequent results, it can be seen that this term plays a decisive role in the late stages of inflation. Therefore, we retain this term in the initial condition \eqref{e53}.

Using the asymptotic conditions of the Bessel function and the initial condition \eqref{e53}, the coefficients $C_1,C_2$ in the general solution \eqref{e50} can be determined as:
\begin{subequations}
	\begin{align}
		C_1&=-\frac{\pi}{\sqrt{k}}\left(\frac{\bar{f}'}{\bar{f}}+ik\right)\frac{\exp \left[i\frac{\pi}{2}(1-\gamma)\right]}{\cos(\pi\gamma)}\label{e54a}\\
		C_2&=-\frac{\pi}{\sqrt{k}}\left(\frac{\bar{f}'}{\bar{f}}+ik\right)\frac{\exp \left[i\frac{\pi}{2}\gamma\right]}{\cos(\pi\gamma)}\label{e54a}
	\end{align}
\end{subequations}
Therefore the suloution of $\varpi$ is 
\begin{align}
	\varpi(k,\eta)=&-\frac{\pi\sqrt{-\eta}}{\bar{f}}\left(\frac{\bar{f}'}{\bar{f}}+ik\right)\left\{\frac{\exp \left[i\frac{\pi}{2}(1-\gamma)\right]}{\cos(\pi\gamma)}J_{\gamma+1/2}(-k\eta)\right.\nonumber\\
	&\left.+\frac{\exp \left[i\frac{\pi}{2}\gamma\right]}{\cos(\pi\gamma)}J_{-\gamma-1/2}(-k\eta)\right\}	\label{e55}
\end{align}	
At the end of inflation, using the super-horizon scale approximation, the asymptotic behavior of the above solution is:
\begin{align}
	\varpi \xrightarrow{-k\eta\to 0}&-\frac{\pi}{\bar{f}\sqrt{k}}\left(\frac{\bar{f}'}{\bar{f}}+ik\right)\times\nonumber\\
	&\left[\mathscr{C}(\gamma)(-k\eta)^{\gamma+1}+\mathscr{C}(-\gamma-1)(-k\eta)^{-\gamma}\right]	\label{e56}
\end{align}
where
\begin{equation}
	\label{e57}
	\mathscr{C}(\gamma):=\frac{\exp\left[i\frac{\pi}{2}(1-\gamma)\right]}{\cos(\pi\gamma)\Gamma\left(\gamma+\frac{3}{2}\right)2^{\gamma+1/2}}
\end{equation}
If only the dominant term is retained, the above expression can be approximated as:
\begin{equation}
	\label{e58}
	\varpi(k,\eta)\to -\frac{\pi}{\bar{f}\sqrt{k}}\left(\frac{\bar{f}'}{\bar{f}}+ik\right)\mathscr{C}(\alpha-1)(-k\eta)^{\alpha}
\end{equation}
where
\begin{equation}
	\label{e59}
	\alpha:=
	\begin{cases}
		\gamma+1&\left(\gamma<-\frac{1}{2}\right)\\
		-\gamma&\left(\gamma>-\frac{1}{2}\right)
	\end{cases}
\end{equation}
We still keep the $\bar{f}'/\bar{f}$ term in equation \eqref{e58} because it is the dominant term in the late stages of inflation, whereas the $ik$ term can be ignored. From the subsequent discussion, it can be seen that this will yield an electromagnetic power spectrum different from that of the conventional inflationary magnetogenesis models.
\section{Power spectrum of electromagnetic field \label{s4} }
\subsection{Power spectrum of electric field\label{s4a}}
To obtain the power spectrum of the electric field, one need to compute the convolution in equation $\eqref{e46a}$. For convenience in calculation, we will adopt some approximations. First, we focus only on the late stages of inflation. At this stage, we assume that all relevant scales have already exceeded the horizon scale. Therefore $|k\eta|\ll 1,|\kappa \eta|\ll 1,\Big||{\bm k}-{\bm \kappa}|\eta\Big|\ll 1$.

Additionally, before the mode leaves the horizon, the ratio of $\Psi_{\bm \kappa}$ to $\delta\phi_{\bm \kappa}$ is proportional to $a/\kappa$ \cite{weinberg2008cosmology}. Even after this mode exits the horizon, the ratio remains small for some time. It begins to increase once reheating starts \cite{dodelson2024modern}. Therefore, we ignore $\Psi$ in the following discussion and focus only on the influence of $\delta\phi$ on the electromagnetic power spectrum.

First, rewrite Equation \eqref{e46a} as:
\begin{equation}
	\label{e61}
	\hat{\varpi}(k,\eta)=\varpi(k,\eta)+\mathbb{I}(k,\eta)
\end{equation}
where
\begin{equation}
	\label{e75}
	\mathbb{I}(k,\eta)=\int \frac{d^3{\bm \kappa}}{(2\pi)^3}\left[-
	\frac{\theta}{\bar{f}}\delta\phi_{\bm \kappa}\right]\varpi({\bm k}-{\bm\kappa},\eta)
\end{equation}

Using the asymptotic behavior of $\delta\phi$ in the late stages of inflation\cite{dodelson2024modern}:
\begin{equation}
	\label{e60}
	\delta\phi_{\kappa}\to \frac{1}{a\sqrt{2\kappa}}e^{-i\kappa\eta}\frac{-i}{\kappa\eta}
\end{equation}
and inserting \eqref{e58} and \eqref{e60} into \eqref{e61}, we have

\begin{align}
	\label{e62}
	\mathbb{I}(k,\eta)\sim&\frac{\theta\pi}{a\bar{f}^2}\mathscr{C}(\alpha-1)\int\frac{d^3{\bm \kappa}}{(2\pi)^3}\frac{-i}{\kappa\eta\sqrt{2\kappa}}e^{-i\kappa\eta}\nonumber\\
	&\left[\frac{\bar{f}'}{\sqrt{|{\bm k}-{\bm \kappa}|}\bar{f}}+i\sqrt{|{\bm k}-{\bm \kappa}|}\right]\cdot\left[-|{\bm k}-{\bm \kappa}|\eta\right]^{\alpha}
\end{align}
For an approximate evaluation of $\mathbb{I}$, we divide the integral in $\mathbb{I}$ into two parts: $\mathbb{I}=\mathbb{I}_1+\mathbb{I}_2$. Where $\mathbb{I}_1$ represents the contribution from the integral over the range $k$ to $\infty$, and $\mathbb{I}_2$ corresponds to the contribution from the integral over the range $0$ to $k$.

For $\mathbb{I}_1$, we apply the approximation $\kappa\gg k$, then $|{\bm k}-{\bm\kappa}|\sim\kappa$, and $\mathbb{I}_1$ can be integrated approximately as:
\begin{align}
	\mathbb{I}_1\sim& \frac{i\theta\mathscr{C}(\alpha-1)}{2\sqrt{2}a\bar{f}^2\pi}(-\eta)^{\alpha-1}\nonumber\\
	&\left[\frac{\bar{f}'}{\bar{f}}k^{\alpha+1}\mathcal{E}_{-\alpha}(ik\eta)+ik^{\alpha+2}\mathcal{E}_{-\alpha-1}(ik\eta)\right]	\label{e63}
\end{align}
where $\mathcal{E}_n(z)$ is exponential integral function and defined as:
\begin{equation}
	\label{e64}
	\mathcal{E}_n(z):=\int_1^{\infty}t^{-n}e^{-zt}dt
\end{equation}
For $\mathbb{I}_2$, we apply the approximation $\kappa\ll k$, then $|{\bm k}-{\bm\kappa}|\sim k$, and one can calculate that $\mathbb{I}_2\sim 0$ which means that $\mathbb{I}\sim\mathbb{I}_1$. Therefore the power spectrum of electric field can be 
written as
\begin{equation}
	\label{e65}
	\mathscr{P}_E\sim \bar{\mathscr{P}}_E+\delta\mathscr{P}_E
\end{equation}
where $\bar{\mathscr{P}}_E$ is the main part of electric field power spectrum:
	\begin{equation}\label{e66}
		\bar{\mathscr{P}}_E=\frac{\bar{f}^2}{8\pi^3}\frac{k^3}{a^4}|\varpi|^2
	\end{equation}
and $\delta\mathscr{P}_E$ represents the contribution of inhomogeneous perturbations to the electric field power spectrum:
\begin{equation}
	\label{e67}
	\delta\mathscr{P}_E=\frac{\bar{f}^2}{8\pi^3}\frac{k^3}{a^4}(\varpi^*\mathbb{I}+\varpi\mathbb{I}^*)
\end{equation}
Insert \eqref{e58} into \eqref{e67} one can get the main part of electric field power spectrum is 
\begin{equation}
	\label{e68}
	\bar{\mathscr{P}}_E=\frac{|\mathscr{C}(\alpha-1)|^2H^4}{8\pi}(-k\eta)^{2\alpha+4}\left[\gamma^2(-k\eta)^{-2}+1\right]
\end{equation}
If the $\bar{f}'/\bar{f}$ term is ignored in the initial conditions \eqref{e53}, it can be verified that the above electric field power spectrum is consistent with the result in the conventional inflationary magnetogenesis model, i.e. when $\alpha=-2~(\gamma=-3~{\rm or}~2)$, it corresponds to a scale-invariant spectrum \cite{2010_subramanian}. However, as mentioned earlier, in the late stage of inflation, the $\bar{f}'/\bar{f}$ term dominates in \eqref{e58}. Therefore, the above electric field power spectrum is:
\begin{equation}
	\label{e69}
	\bar{\mathscr{P}}_E\sim\frac{|\mathscr{C}(\alpha-1)|^2H^4\gamma^2}{8\pi}(-k\eta)^{2\alpha+2}
\end{equation}
and the index of spectrum of electric field is 
\begin{equation}
	\label{e70}
	\bar{n}_E=\frac{d\ln \bar{\mathscr{P}}_E}{d\ln k}=2\alpha+2
\end{equation}
which means that the electric field is scale-invariant only when $\alpha=-1~(\gamma=-2~{\rm or}~1)$. 

Insert \eqref{e58} and \eqref{e63} into \eqref{e67} one can obtain the contribution of inhomogeneous perturbations to the electric field power spectrum:
\begin{equation}
	\label{e71}
	\delta \mathscr{P}_E\sim \frac{k^2|\mathscr{C}(\alpha-1)|^2H^5\gamma^2}{8\sqrt{2}\pi^3}\frac{\theta}{\bar{f}}(-k\eta)^{2\alpha+2}
	\Im[\mathscr{E}_{\alpha}(-k\eta)]
\end{equation}
where
\begin{equation}
	\label{e72}
	\mathscr{E}_{\alpha}(z):=\mathcal{E}_{-\alpha}(-iz)-iz\mathcal{E}_{-\alpha-1}(-iz)
\end{equation}
The asymptotic behavior of function $\mathscr{E}$ as $z\to 0$ depends on the value of $\alpha$:
\begin{equation}
	\label{e73}
	\mathscr{E}_{\alpha}(z)\xrightarrow{z\to 0}
	\begin{cases}
		\frac{(-i)^{-\alpha}}{z^{1+\alpha}}\left[i\Gamma(1+\alpha)+\frac{1}{\gamma}\Gamma(2+\alpha)\right]&(\alpha>-1)\\
		-\ln|z|-i\frac{\pi\gamma-2}{2\gamma}&(\alpha=-1)\\
		-\frac{1}{1+\alpha}-i\frac{z}{2+\alpha}&(\alpha<-1)
	\end{cases}
\end{equation}
For convenience in discussion, we define the relative impact of inhomogeneous perturbations on the electric field power spectrum as:
\begin{equation}
	\label{e74}
	\Delta_E:=\frac{\delta\mathscr{P}_E}{\bar{\mathscr{P}}_E}=\frac{k^2H}{\sqrt{2}\pi^2}\frac{\theta}{\bar{f}}\Im [\mathscr{E}_{\alpha}(-k\eta)]
\end{equation}
\subsection{Power spectrum of magnetic field}
To calculate the magnetic field power spectrum, the mode function $\mathcal{A}$ should be computed first. For this purpose, rewrite \eqref{e41} as: 
\begin{equation}
	\label{e76}
	(a\mathcal{A})'=\varpi+2\mathbb{I}~\Rightarrow~a\mathcal{A}=\int d\tau (\varpi+2\mathbb{I})
\end{equation}
Insert \eqref{e76} into \eqref{e46b}:
\begin{equation}
	\label{e77}
	\hat{\mathcal{A}}_i=k_i\left[\int d\tau (\varpi+2\mathbb{I})+\mathbb{J}\right]
\end{equation}
where
\begin{equation}
	\label{e78}
	\mathbb{J}:=\int\frac{d^3{\bm \kappa}}{(2\pi)^3}\frac{\theta}{\bar{f}}\delta\phi_{\bm\kappa}\left[\int d\tau\varpi({\bm k}-{\bm\kappa},\tau)\right]
\end{equation}
Therefore, to calculate the magnetic field power spectrum, the following two time integrals need to be evaluated:
\begin{equation}
	\label{e79}
	\mathscr{K}:=\int_{-\infty}^{\eta} d\tau\varpi,\hskip 10pt \mathscr{R}:=\int_{-\infty}^{\eta} d\tau\mathbb{I}
\end{equation}
It is worth noting that when calculating $\mathscr{K}$, one should first compute the time integral of equation \eqref{e55} and then take the super-horizon limit, rather than directly computing the time integral of equation \eqref{e58} since that \eqref{e58} only describes the asymptotic behavior during the late stage of inflation.
Therefore from \eqref{e55}, one can obtain the time integral of $\varpi$, if only the leading term is considered then 
\begin{widetext} 
	\begin{align}
		\mathscr{K}\sim-\frac{\pi}{\sqrt{k}\bar{f}}(-k\eta)^{\gamma}&\left\{
		\frac{\exp\left[i\frac{\pi}{2}(1-\gamma)\right]}{\cos(\pi\gamma)}
		\left[\frac{-2^{-1/2-\gamma}\sqrt{\pi}\gamma}{\Gamma(1+\gamma)}-\frac{-2^{1/2-\gamma}i}{\Gamma(1/2+\gamma)}\right]\right.\nonumber\\
		&\left.+\frac{\exp\left(i\frac{\pi}{2}\gamma\right)}{\cos(\pi\gamma)}(-k\eta)^{-2\gamma}
		\left[\frac{(-1)^{\gamma-1}2^{\gamma-1/2}\Gamma(\gamma+1/2)}{\pi}+\frac{ik\eta2^{-1/2+\gamma}}{\Gamma(3/2-\gamma)}\right]
		\right\} 
		\label{e80}
	\end{align}
\end{widetext}
Similar to the calculation of $\mathscr{K}$, when computing $\mathscr{R}$, one should first perform the time integral and then take the super-horizon approximation. For this purpose, $\mathscr{R}$ is rewritten as:
\begin{equation}
	\label{e81}
	\mathscr{R}=-\frac{1}{k}\int_{\infty}^{-k\eta}\mathbb{I}(-k\tau)d(-k\tau)
\end{equation}

To compute $\mathscr{R}$, it is necessary to first perform the integration of $\mathbb{I}$ with respect to $\tau$, and then take the superhorizon limit. Therefore, equation \eqref{e62} cannot be integrated directly. 

Insert \eqref{e55} into \eqref{e75}, and 
use the complete expression for the inhomogeneous perturbations of the inflation field \cite{dodelson2024modern}:
\begin{equation}
	\label{e82}
	\delta\phi_{\kappa}=\frac{1}{a\sqrt{2\kappa}}e^{-i\kappa\tau}\left(1-\frac{i}{\kappa\tau}\right)
\end{equation}
 the full expression for $\mathbb{I}$ can be obtained. Similar to the calculation of equation \eqref{e62}, we divide the integral into two parts: $\mathbb{I}=\mathbb{I}_1+\mathbb{I}_2$. 

$\mathbb{I}_1$ represents the contribution from the integral over the range $k$ to $\infty$, therefore we apply the approximation 
$\kappa\gg k$; $\mathbb{I}_2$ represents the contribution from the integral over the range $0$ to $k$, therefore we apply the approximation $\kappa\ll k$,then:
\begin{widetext}
	\begin{align}
		\mathbb{I}_1\sim&\frac{\theta\sqrt{-\tau}}{2\sqrt{2}a\pi\bar{f}^2}\int_k^{\infty}\kappa^{3/2}e^{i\kappa\tau}\left(1-\frac{i}{\kappa\tau}\right)\left(\frac{\bar{f}'}{\bar{f}}+i\kappa\right)
		\left\{\frac{\exp\left[i\frac{\pi}{2}(1-\gamma)\right]}{\cos(\pi\gamma)}J_{\gamma+\frac{1}{2}}(-\kappa\tau)
		+\frac{\exp\left(i\frac{\pi}{2}\gamma\right)}{\cos(\pi\gamma)}J_{-\gamma-\frac{1}{2}}(-\kappa\tau)\right\}d\kappa\label{ea1}\\
		\mathbb{I}_2\sim&\frac{\theta\sqrt{-\tau}}{2\sqrt{2}a\pi\bar{f}^2}\left(\frac{\bar{f}'}{\bar{f}}+ik\right)
		\left\{\frac{\exp\left[i\frac{\pi}{2}(1-\gamma)\right]}{\cos(\pi\gamma)}J_{\gamma+\frac{1}{2}}(-k\tau)
		+\frac{\exp\left(i\frac{\pi}{2}\gamma\right)}{\cos(\pi\gamma)}J_{-\gamma-\frac{1}{2}}(-k\tau)\right\}
		\int_0^k\kappa^{3/2}e^{-i\kappa\tau}\left(1-\frac{i}{\kappa\tau}\right)d\kappa\label{ea2}
	\end{align}
\end{widetext}

Divide the $\mathscr{R}$ into two parts: $\mathscr{R}=\mathscr{R}_1+\mathscr{R}_2$, where $\mathscr{R}_1$ is the integral from $\infty$ to $1$, and $\mathscr{R}_2$ is the integral from $1$ to $-k\eta$. Considering that the limit $-k\eta\ll1$ is taken after the integration, for $\mathscr{R}_1$ one can adopt the approximation $-k\tau\gg 1$, and for $\mathscr{R}_2$ the approximation $-k\tau\ll1$.

To obtain the $\mathscr{R}_1$, the approximation $-k\tau\gg 1$ lead to:
\begin{equation}
	\label{ea3}
	\mathbb{I}_1\sim\frac{i\theta_0 H}{2f_0^2\pi^{3/2}}(-\tau)^{\beta-2\gamma-3/2}\int_{-k\tau}^{\infty}z^2e^{2iz}dz
\end{equation}
where $z:=-\kappa\tau$ and we set $\bar{f}=f_0\tau^{\gamma},~\theta=\theta_0(-\tau)^{\beta}$. Notice that the integral of $z$ in
\eqref{ea3} is $\mathcal{O}[(-\tau)^3]$, then if $\beta-2\gamma-\frac{3}{2}>-3$, $\mathbb{I}_1$ will diverges for $-k\tau\gg 1$, and the perturbative approach will break down. Therefore we only consider the case where $\beta-2\gamma<-3/2$. In this case, $\mathbb{I}_1\sim 0$.

In the calculation of $\mathbb{I}_2$, since $\kappa\ll k$, $-k\tau\gg 1$ does not necessarily imply $-\kappa\tau\gg 1$. However, to ensure that all relevant scales satisfy $-k\tau\gg 1$, we actually require $-\tau\gg 1$, which in turn implies $-\kappa\tau\gg 1$. Therefore, in calculating $\mathbb{I}_2$, we adopt the approximation $-k\tau\gg-\kappa\tau\gg 1$, then
\begin{equation}
	\label{ea4}
	\mathbb{I}_2\sim-\frac{\theta k^2}{2\bar{f}^2a\pi^{3/2}\tau}\exp\left(-2ik\tau\right)
\end{equation}
under the condition $\beta-2\gamma<-3/2$, $\mathbb{I}_2\sim 0$ when $-k\tau\gg 1$, therefore $\mathscr{R}_1\sim 0$.

In the calculation of $\mathscr{R}_2$, the approximation $-k\tau\ll 1$ should be used. In this case, as discussed in subsection \ref{s4a}, $\mathbb{I}\sim\mathbb{I}_1$, where the expression for $\mathbb{I}_1$ is given in equation \eqref{e63} with the time variable is $\tau$.
Therefore 
\begin{equation}
	\label{ea5}
	\mathscr{R}_2=-\frac{i\theta_0\mathscr{C}(\alpha-1)H\gamma}{2\sqrt{2}f_0^2\pi k^{\beta-2\gamma-1}}\int_1^{-k\eta}z^{\alpha+\beta-2\gamma-1}\mathscr{E}_{\alpha}(z)dz
\end{equation}
If only the leading term is considered, the resulting expression for $\mathscr{R}\sim\mathscr{R}_2$ is:
 \begin{equation}
 	\label{e83}
 	\mathscr{R}\sim\frac{i\theta\mathscr{C}(\alpha-1)\gamma}{2\sqrt{2}\bar{f}^2\pi a(\beta-2\gamma-1)}\mathscr{F}_{\alpha}(-k\eta)
 \end{equation}
 where 
 \begin{align}
 	\label{e89}
 	&\mathscr{F}_{\alpha}(-k\eta):=\nonumber\\
 	&
 	\begin{cases}
 		(-i)^{1-\alpha}\Gamma(1+\alpha)-(-i)^{-\alpha}\Gamma(2+\alpha)\gamma^{-1}&(\alpha>-1)\\
 		(-k\eta)^{\alpha+1}\frac{\beta-2\gamma-1}{(1+\alpha)(\alpha+\beta-2\gamma)}&(\alpha<-1)\\
 		\ln(-k\eta)+i\left(\gamma^{-1}-\frac{\pi}{2}\right)&(\alpha=-1)
 	\end{cases}
 \end{align}
 
 At last, insert \eqref{e60} and \eqref{e80} into \eqref{e78}, and divide the integral into two parts as in calculate of $\mathbb{I}$, one can get:
 \begin{equation}
 	\label{e84}
 	\mathbb{J}\sim\frac{ik^2\theta\mathscr{D}(\gamma)}{2\sqrt{2}\pi^2a\bar{f}^2}(-k\eta)^{-\gamma-1}\mathcal{E}_{\gamma}(ik\eta)
 \end{equation}
 where 
 \begin{equation}
 	\label{e85}
 	\mathscr{D}(\gamma)=(-1)^{\gamma}2^{\gamma-1/2}\Gamma\left(\gamma+\frac{1}{2}\right)\frac{\exp\left(i\frac{\pi}{2}\gamma\right)}{\cos(\pi\gamma)}
 \end{equation}
 
Similar to the power spectrum of electric field, the magnetic field power spectrum can be written as
 \begin{equation}
 	\label{e86}
 	\mathscr{P}_B\sim \bar{\mathscr{P}}_B+\delta\mathscr{P}_B
 \end{equation}
 where $\bar{\mathscr{P}}_B$is the main part of magnetic field power spectrum:
 \begin{equation}
 	\label{e87}
 	\bar{\mathscr{P}}_B=\frac{\bar{f}^2k^5}{8\pi^3a^4}\left|\mathscr{K}\right|^2
 \end{equation}
 and $\delta\mathscr{P}_B$is the contribution of inhomogeneous perturbations to the magnetic field power spectrum:
 \begin{equation}
 	\label{e88}
 	\delta\mathscr{P}_B=\frac{\bar{f}^2k^5}{8\pi^3a^4}\left[\mathscr{K}^*\left(2\mathscr{R}+\mathbb{J}\right)
 	+\mathscr{K}\left(2\mathscr{R}^*+\mathbb{J}^*\right)\right]
 \end{equation}
It can be verified that if the $\bar{f}'/\bar{f}$ term is neglected in the initial conditions, Equation \eqref{e87} yields the standard result of the conventional inflationary magnetogenesis model, namely that a scale-invariant spectrum is obtained when $\gamma=-2$ or $\gamma=3$ \cite{2010_subramanian}. However, as in the discussion of electric field, we retain this term in the initial conditions, then
\begin{equation}
	\label{e90}
	\bar{\mathscr{P}}_B=\frac{H^4|\mathscr{D}(\gamma)|^2}{8\pi^3}(-k\eta)^{4-2\gamma}
\end{equation} 

	In the inflationary magnetogenesis model, there exists a strong coupling problem\cite{2009_demozzi}. The effective coupling constant between the electromagnetic field and matter is given by $e_{\rm eff}=e/f$. Therefore, to ensure that $e_{\rm eff}=e$ at the end of inflation, it is required that $f=1$ at the end of inflation. If $f$ increases over time during inflation, then it would be very small at the beginning of inflation, which leads to a large $e_{\rm eff}$, thereby causing the strong coupling problem. To avoid this problem, $f$ should be a decreasing function during inflation, which requires $\gamma>0$. In the following, we consider only this case, then the index of spectrum of magnetic field is 

\begin{equation}
	\label{e91}
	\bar{n}_B=\frac{d\ln \bar{\mathscr{P}}_B}{d\ln k}=4-2\gamma
\end{equation}
which means that the scale-invariant spectrum of magnetic can be obtained when $\gamma=2$. At this case, the index of spectrum of 
electric field is $n_E=-2$, and as a result, the backreaction problem arises.

Substituting the \eqref{e80},\eqref{e83} and \eqref{e84} into Equations \eqref{e88}, we obtain the contribution of inhomogeneous 
perturbations to the magnetic field power spectrum
\begin{equation}
	\label{e92}
	\delta\mathscr{P}_B\sim \frac{-|\mathscr{D}(\gamma)|^2\theta H^5}{8\sqrt{2}\pi^5\bar{f}\sqrt{k}}(-k\eta)^{4-2\gamma}
	\Im\left[\mathcal{E}_{\gamma}(ik\eta)\right]
\end{equation}
The relative impact of inhomogeneous perturbations on the magnetic field power spectrum is:
\begin{equation}
	\label{e93}
	\Delta_B:=\frac{\delta\mathscr{P}_B}{\bar{\mathscr{P}}_B}=-\frac{H\theta}{\sqrt{2k}\pi^2\bar{f}}\Im\left[\mathcal{E}_{\gamma}(ik\eta)\right]
\end{equation}

\subsection{Backreaction problem}
As outlined in the previous subsection, to circumvent the strong coupling problem, we restrict our attention to the case $\gamma>0$. However, under this condition, a potential backreaction issue may emerge—specifically, the energy density of the generated electromagnetic fields could become comparable to or even exceed that of the inflaton, thereby undermining the validity of the perturbative framework employed. This subsection is devoted to examining the constraints on the parameter $\gamma$ imposed by the requirement of avoiding such backreaction effects.

In the case of $\gamma>0$ , the main part of power spectrum of the electromagnetic field is given by: 
\begin{align}
	\bar{\mathscr{P}}_{\rm E}&\sim \frac{|\mathscr{C}(-\gamma-1)\gamma|^2H^4}{8\pi}\left(\frac{k}{aH}\right)^{2-2\gamma}\label{ea1}\\
	\bar{\mathscr{P}}_{\rm B}&\sim \frac{|\mathscr{D}(\gamma)/\pi|^2H^4}{8\pi}\left(\frac{k}{aH}\right)^{4-2\gamma}\label{ea2}
\end{align}
To avoid the backreaction problem, the following condition should be satisfied \cite{2008_martin}:
\begin{equation}
	\label{ea3}
	\bar{\mathscr{P}}_{\rm E}\Big|_{\rm inf}+\bar{\mathscr{P}}_{\rm B}\Big|_{\rm inf}<\rho_{\rm inf}=\frac{3}{8\pi}H^2_{\rm inf}m_{pl}^2
\end{equation}
where "inf" denotes the value at the end of inflation and $m_{pl}=1/\sqrt{G}$ is Pkanck mass.

Insert \eqref{ea1} and \eqref{ea2} into \eqref{ea3}:
\begin{align}
	\frac{H^4_{\rm inf}}{8\pi}&\left[|\mathscr{C}(-\gamma-1)\gamma|^2\left(\frac{a_0H_0}{a_{\rm inf}H_{\rm inf}}\right)^{2-2\gamma}\left(\frac{k}{a_0H_0}\right)^{2-2\gamma}\right.\nonumber\\
	&\left.|\mathscr{D}(\gamma)/\pi|^2\left(\frac{a_0H_0}{a_{\rm inf}H_{\rm inf}}\right)^{4-2\gamma}\left(\frac{k}{a_0H_0}\right)^{4-2\gamma}\right]<\rho_{\rm inf}\label{ea4}
\end{align}
where "0" denotes the value today. If we assume that the energy density at the end of inflation is approximately equal to the energy density when the scales of interest today exited the horizon, then \cite{2006_martin}:
\begin{equation}
	\label{ea5}
	\frac{a_0H_0}{a_{\rm inf}H_{\rm inf}}\sim 1.517\times 10^{-29}\frac{h}{R}
\end{equation}
Here, $R$ is a parameter related to reheating. If we further assume that reheating is instantaneous, then\cite{2006_martin,2008_martin}:
\begin{equation}
	\label{ea6}
	\ln R\sim \frac{1}{4}\ln\left(\frac{\rho_{\rm inf}}{m_{pl}^4}\right)
\end{equation}

\begin{figure}[htbp]
	\includegraphics[width=\columnwidth]{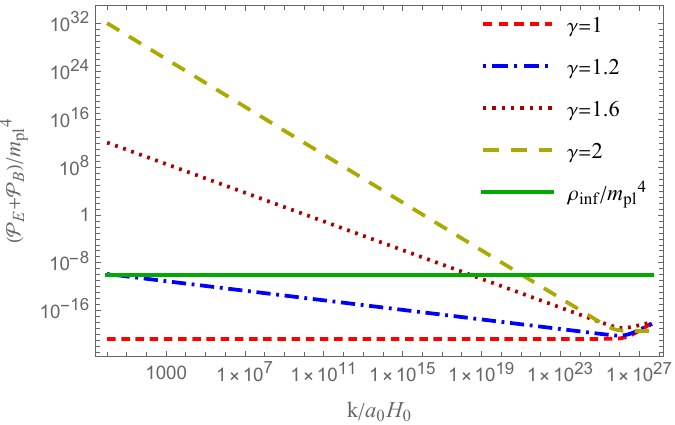}
	\caption{The electromagnetic field spectrum at the end of inflation for different $\gamma$. The green solid line indicates the inflaton energy density which is $10^{-10}m_{pl}^4$.}
	\label{fa1}
\end{figure}

Fig.\ref{fa1} shows the curves of $\bar{\mathscr{P}}_{\rm E}+\bar{\mathscr{P}}_{\rm B}$ versus $k/a_0H_0$ at the end of inflation for different values of $\gamma$.  The green solid line indicates the inflaton energy density. Here, $\rho_{\rm inf}\sim10^{-10}m_{pl}^4$ (GUT energy scale) is the upper limit of the inflaton energy density \cite{2006_martin}. 

Since the minimal number of e-folds required for inflation depends on the energy density of the inflaton, it suffices to consider only those modes that exited the horizon during inflation. As shown in the Fig.\ref{fa1}, for $\gamma < 1.2$, the backreaction problem can always be avoided. This range includes the case $\gamma = 1$, where the electric field spectrum is scale-invariant. However, for $\gamma > 1.2$, the electromagnetic power spectrum exceeds the inflaton energy density at large scales, leading to a backreaction problem. This includes the case $\gamma = 2$, where the magnetic field spectrum is scale-invariant. In other words, if inflation proceeds with the maximum allowed energy density, a scale-invariant magnetic field spectrum will inevitably lead to backreaction issues.

\begin{figure}[htbp]
	\includegraphics[width=\columnwidth]{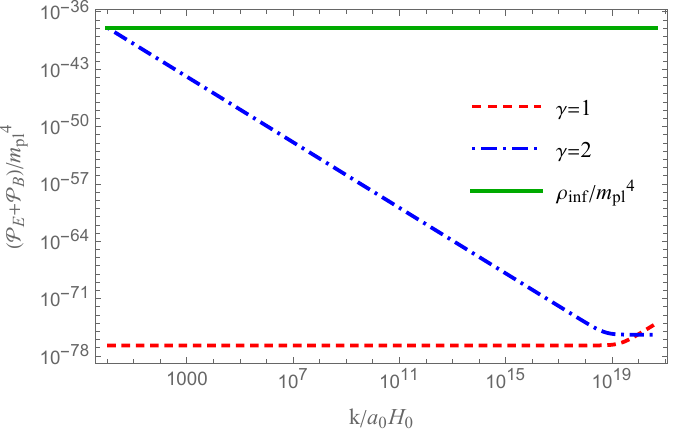}
	\caption{The electromagnetic field spectrum at the end of inflation for different $\gamma$. The green solid line indicates the inflaton energy density which is $10^{-38}m_{pl}^4$.}
	\label{fa2}
\end{figure}

Conversely, if one requires a scale-invariant primordial magnetic field spectrum, then avoiding the backreaction problem imposes constraints on the upper limit of the inflaton energy density. As shown in Fig.\ref{fa2}, to avoid backreaction for $\gamma = 2$, one must have $\rho_{\rm inf}<10^{-38}m^4_{pl}$. Notably, the lower bound on the inflaton energy density is $10^{-85}m^4_{pl}$, set by the Big Bang Nucleosynthesis (BBN) energy scale. Therefore, it is indeed possible to construct an inflationary magnetogenesis scenario that avoids the backreaction problem while producing a scale-invariant magnetic field spectrum.

\section{Inflationary Models\label{sa1}}

\subsection{Deviation of spectrum index}
From equations \eqref{e74} and \eqref{e93}, it can be seen that the effect of inhomogeneous scalar perturbations on the electromagnetic spectrum is proportional to $\theta/\bar{f}$, and $\theta/\bar{f}$ depends on the specific inflationary model. Noting that
\begin{equation}
	\label{e94}
	\bar{f}'=\frac{d\bar{f}}{d\bar{\phi}}\bar{\phi}'=\theta\bar{\phi}'~\Rightarrow~\frac{\theta}{\bar{f}}=\frac{\bar{f}'}{\bar{f}}\left(\bar{\phi}'\right)^{-1}
\end{equation}
Insert \eqref{e94} into \eqref{e74} and \eqref{e93} and using $\bar{f}'/\bar{f}=\gamma\eta^{-1}$, one can get
\begin{subequations}
	\begin{align}
		\Delta_E=&\frac{k^2H\gamma}{\sqrt{2}\pi^2}\eta^{-1}\Im\left[\mathscr{E}_{-\gamma}(-k\eta)\right] \left(\bar{\phi}'\right)^{-1}\label{e95a}\\
		\Delta_B=&\frac{-H\gamma}{\sqrt{2k}\pi^2}\eta^{-1}\Im\left[\mathcal{E}_{\gamma}(ik\eta)\right]\left(\bar{\phi}'\right)^{-1}\label{e95b}
	\end{align}
\end{subequations}
where we only consider the case of $\gamma>0$. The characteristics of the inflationary model are reflected in the $\left(\bar{\phi}'\right)^{-1}$ factor present in the above two equations, while the ratio between $\Delta_E$ and $\Delta_B$ is independent of the specific inflationary model:
\begin{equation}
	\label{e96}
	\frac{\Delta_E}{\Delta_B}=-k^{5/2}\frac{\Im\left[\mathscr{E}_{-\gamma}(-k\eta)\right]}{\Im\left[\mathcal{E}_{\gamma}(ik\eta)\right]}
\end{equation}
We focus only on the cases where the electric or magnetic field has a scale-invariant spectrum, i.e. $\gamma=1$ for scale-invariant spectrm of electric field and $\gamma=2$ for scale-invariant spectrum of magnetic field, then 
\begin{align}
	\Im\left[\mathscr{E}_{-\gamma}(-k\eta)\right]=&
	\begin{cases}
		\frac{\pi}{2}&(\gamma=1)\\
		-k\eta&(\gamma=2)
	\end{cases}\label{e97}\\
	\Im\left[\mathcal{E}_{\gamma}(ik\eta)\right]=&
	\begin{cases}
		\frac{\pi}{2}&(\gamma=1)\\
		-\ln(-k\eta)(-k\eta)&(\gamma=2)
	\end{cases}\label{e98}
\end{align}
Therefore
\begin{equation}
	\label{e99}
	\frac{\Delta_E}{\Delta_B}=
	\begin{cases}
		-k^{5/2}&(\gamma=1)\\
		\frac{k^{5/2}}{\ln(-k\eta)}\to 0&(\gamma=2)
	\end{cases}
\end{equation}
This result indicates that if the electric field has a scale-invariant spectrum ($\gamma=1$), the influence of inhomogeneous perturbations of the inflaton field on the electric and magnetic spectra with a fixed scale is proportional. However, if the magnetic field has a scale-invariant spectrum ($\gamma=2$), the influence of the inflaton’s inhomogeneous perturbations on the magnetic field becomes increasingly dominant compared to their effect on the electric field spectrum.

To better discuss the influence of inhomogeneous perturbations on the spectrum of electromagnetic field, let's now turn our attention to the spectrum index of the electromagnetic field. In fact, the spectrum index of the electric and magnetic fields defined by equations \eqref{e70} and \eqref{e91} represent only their leading components. The complete spectral indices should be:
\begin{align}
	n_{E/B}:=&\frac{d\ln\mathscr{P}_{E/B}}{d\ln k}=\frac{d}{d\ln k}\ln\left[\bar{\mathscr{P}}_{E/B}(1+\Delta_{E/B})\right]\nonumber\\
	=&\bar{n}_{E/B}+\delta n_{E/B}
	\label{e100}
\end{align}
where $\delta n_{E/B}$ are the the deviation of the spectrum index from its leading part due to the inhomogeneous perturbations of the inflaton:
\begin{equation}
	\label{e101}
	\delta n_{E/B}:=\frac{d\ln(1+\Delta_{E/B})}{d\ln k}\sim k\frac{d\Delta_{E/B}}{dk}
\end{equation}
When electric field has scale-invariant spectrum ($\gamma=1$), $\bar{n}_E=0,~\bar{n}_B=2$, then
\begin{subequations}
	\begin{align}
		\delta n_E=&-\frac{k^2aH^2}{\sqrt{2}\pi\bar{\phi}'}\label{e102a}\\
		\delta n_B=&-\frac{aH^2}{4\sqrt{2k}\pi\bar{\phi}'}\label{e102b}
	\end{align}
\end{subequations}
Under the slow-roll approximation
\begin{equation}
	\label{e103}
	\bar{\phi}'\sim-\frac{V_{\phi}}{3H}\Rightarrow\frac{aH^2}{\bar{\phi}'}\sim-\frac{3H^3}{V_{\phi}}
\end{equation}
and then
\begin{equation}
	\label{e104}
	\delta n_E\sim\frac{3H^3k^2}{\sqrt{2}\pi V_{\phi}},\hskip 10pt \delta n_B\sim \frac{3H^3}{4\sqrt{2k}\pi V_{\phi}}
\end{equation}
From this result, it can be seen that the deviation of spectrum index depends on the sign of $V_{\phi}$. For large field inflation such as chaotic inflation, $V_{\phi}>0$, the spectrum index will increase which means that the inhomogeneous perturbations of the inflaton field will cause the electromagnetic power spectrum to shift towards the blue end. While for small-field inflation ,natural inflation etc, the electromagnetic power spectrum will shift towards the red end. As the inflation proceeds, the value of $|V_{\phi}|$ gradually increases, which leads to a decrease in the degree of deviation of the spectrum index of both the electric and magnetic fields.

If we consider a fixed moment in time, it can be seen that as the scale increases, the deviation of the electric field spectrum index decreases, while the deviation of the magnetic field spectrum index increases. This implies that when the electric field exhibits a scale-invariant spectrum, the inhomogeneous perturbations of the inflaton have a smaller effect on the electric field at large scales and a greater effect at small scales, whereas the opposite is true for the magnetic field.

On the other hand, if magnetic field has scale-invariant spectrum $(\gamma=2)$, $\bar{n}_E=-2,~\bar{n}_B=0$, then
\begin{equation}
	\label{e105}
	\delta n_E\sim\frac{18k^3H^2}{\sqrt{2}\pi^2V_{\phi}},\hskip 10pt \delta n_B\sim\frac{3H^2}{\sqrt{2}\pi^2V_{\phi}}\ln(-k\eta)\sqrt{k}
\end{equation}
In this case, the effect of the inhomogeneous perturbations of the inflaton on the spectrum index of the electric field is similar to the situation with $\gamma=1$. However, since $\ln(-k\eta)<0$, the effect on the spectrum index of the magnetic field is opposite to that in the $\gamma=1$ case — that is, the magnetic field spectrum experiences a redshift when $V_{\phi}>0$, and a blueshift when $V_{\phi}<0$.

As inflation progresses, the deviation of the electric field spectrum index decreases, similar to the case with $\gamma = 1$. In contrast, the deviation of the magnetic field spectrum index depends on the evolution of $V_{\phi}$ and $\ln (-k\eta)$.

In the following, we take the exponential potential as an example to discuss the behavior of $\Delta_{E/B}$ and $\delta n_{E/B}$.

In the case of an exponential potential:
\begin{equation}
	\label{e106}
	V(\phi)\propto \exp(-\lambda\phi)\Rightarrow V_{\phi}\sim-3H\bar{\phi}'\propto(-\eta)^{2\epsilon}
\end{equation}
where $\epsilon\ll1$ is slow-roll parameter.

Insert \eqref{e106} int \eqref{e95a} and \eqref{e95b} one can get:
\begin{subequations}
	\begin{align}
		\Delta_E\propto& \gamma(-k\eta)^{2\epsilon-1}{\rm Im}\left[\mathscr{E}_{-\gamma}(-k\eta)\right]\label{ea7a}\\
		\Delta_B\propto&-\gamma(-k\eta)^{2\epsilon-1}{\rm Im}\left[\mathcal{E}_{\gamma}(ik\eta)\right]\label{ea7b}
	\end{align}
\end{subequations}
\begin{figure}[htbp]
	\includegraphics[width=\columnwidth]{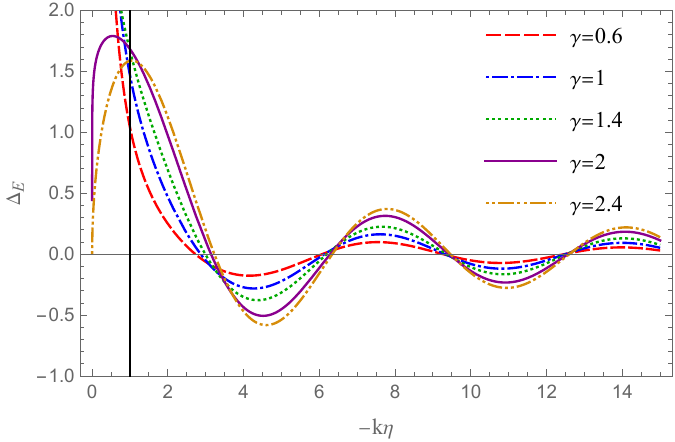}
	\caption{The evolution of $\Delta_E$ for different values of $\gamma$ , where $\epsilon=0.05$}
	\label{fa3}
\end{figure}
\begin{figure}[htbp]
	\includegraphics[width=\columnwidth]{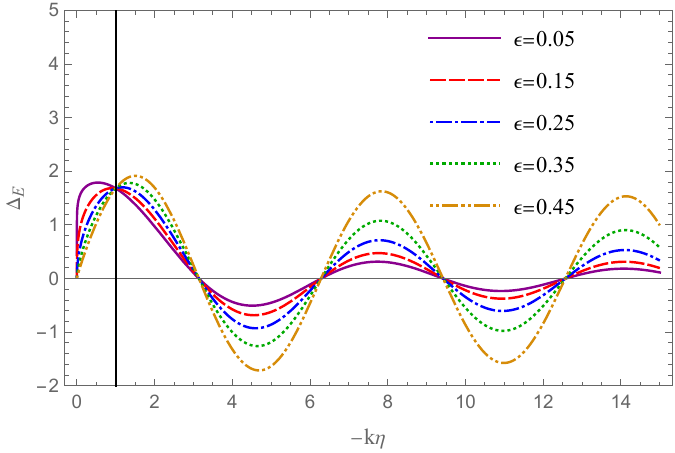}
	\caption{The evolution of $\Delta_E$ for different values of $\epsilon$, where $\gamma=2$.}
	\label{fa4}
\end{figure}

Fig.\ref{fa3} and Fig.\ref{fa4} show the time evolution of $\Delta_E$. In Fig.\ref{fa3}, $\epsilon = 0.05$ and $\gamma$ ranges from 0.6 to 2.4. In Fig.\ref{fa4}, $\gamma = 2$ and $\epsilon$ varies from 0.05 to 0.45. From the figures, it can be seen that the perturbation of the electric field spectrum due to inhomogeneous fluctuations oscillates around $\bar{\mathscr{P}}_E$. As inflation proceeds, the amplitude of these oscillations increases.

Fig.\ref{fa3} shows that a larger $\gamma$ leads to a greater oscillation amplitude. For the case $\gamma \geq 2$, the influence of inhomogeneous perturbations on the electric field spectrum rapidly diminishes once the mode exits the horizon (indicated by the vertical solid line in the figure). This behavior is consistent with the result given in \eqref{e99}.

Furthermore, as shown in Fig.\ref{fa4}, the amplitude of $\Delta_E$ oscillations increases with the slow-roll parameter $\epsilon$.

\begin{figure*}[hbtp]
		\centering
		\begin{subfigure}{0.45\textwidth}
			\centering
			\includegraphics[width=\textwidth]{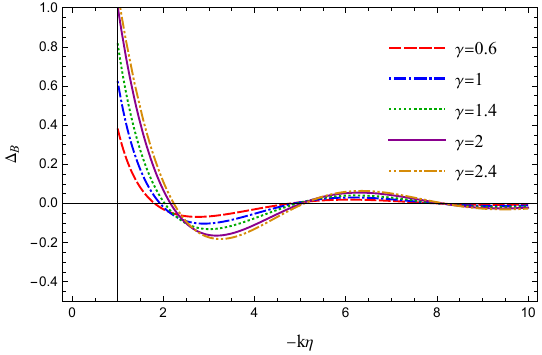}
			\caption{Before horizon exit.}
		\end{subfigure}
		\hfill
		\begin{subfigure}{0.45\textwidth}
			\centering
			\includegraphics[width=\textwidth]{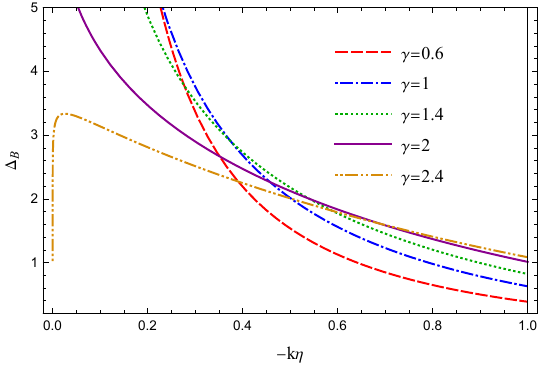}
			\caption{After horizon exit.}
		\end{subfigure}
		\caption{The evolution of $\Delta_B$ for different values of $\gamma$ , where $\epsilon=0.05$}
		\label{fa5}
\end{figure*}
\begin{figure*}[hbtp]
	\centering
	\begin{subfigure}{0.45\textwidth}
		\centering
		\includegraphics[width=\textwidth]{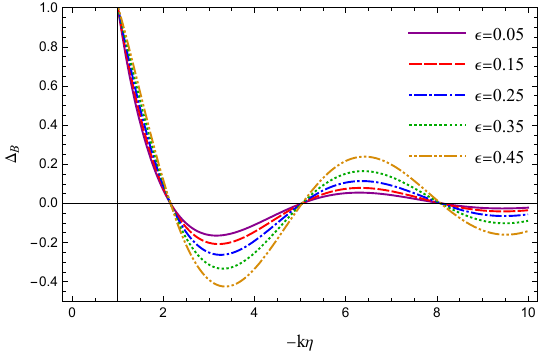}
		\caption{Before horizon exit.}
	\end{subfigure}
	\hfill
	\begin{subfigure}{0.45\textwidth}
		\centering
		\includegraphics[width=\textwidth]{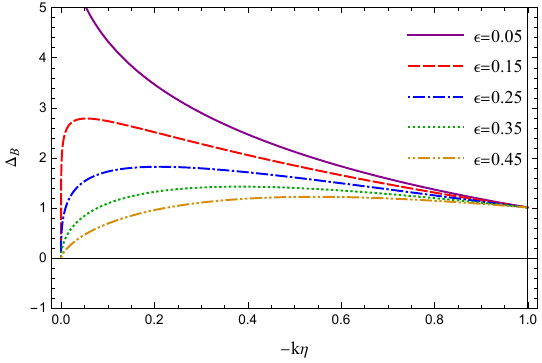}
		\caption{After horizon exit.}
	\end{subfigure}
	\caption{The evolution of $\Delta_B$ for different values of $\epsilon$, where $\gamma=2$.}
	\label{fa6}
\end{figure*}

Fig.\ref{fa5} and Fig.\ref{fa6} show the time evolution of $\Delta_B$. Fig.\ref{fa5}(a) and Fig.\ref{fa6}(a) present the evolution before horizon exit. Similar to $\Delta_E$, the influence of inhomogeneous perturbations on the magnetic field spectrum also oscillates around $\bar{\mathscr{P}}_B$. The larger the values of $\gamma$ and $\epsilon$, the larger the oscillation amplitude.

Fig.\ref{fa5}(b) and Fig.\ref{fa6}(b) depict the evolution of $\Delta_B$ after horizon exit. As shown in Fig.\ref{fa5}(b), for $\gamma \leq 2$, $\Delta_B$ gradually increases, whereas for $\gamma > 2$, $\Delta_B$ rapidly decreases as $-k\eta$ approaches zero. Fig.\ref{fa6}(b) further demonstrates that if the slow-roll parameter $\epsilon$ is not sufficiently small, $\Delta_B$ also rapidly decreases as $-k\eta \rightarrow 0$.

\begin{figure}[htbp]
	\includegraphics[width=\columnwidth]{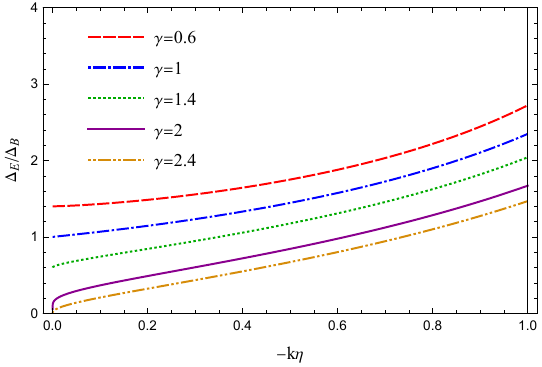}
	\caption{The evolution of $\Delta_E$ for different values of $\gamma$.}
	\label{fa7}
\end{figure}

Fig.\ref{fa7} shows the evolution of $\Delta_E / \Delta_B$ after horizon exit. As seen in the figure, when $\gamma \geq 2$, although both $\Delta_E$ and $\Delta_B$ rapidly decay to zero, the ratio $\Delta_E / \Delta_B$ also tends to zero. This indicates that $\Delta_E$ decays faster than $\Delta_B$, which is consistent with the conclusion drawn in \eqref{e99}.

Insert \eqref{e106} into \eqref{e105} one can get the deviation of spectrum index of magnetic field:
\begin{equation}
	\label{e107}
	\delta n_B\propto-(-k\eta)^{2\epsilon}\ln(-k\eta)
\end{equation}
\begin{figure}[htbp]
	\includegraphics[width=\columnwidth]{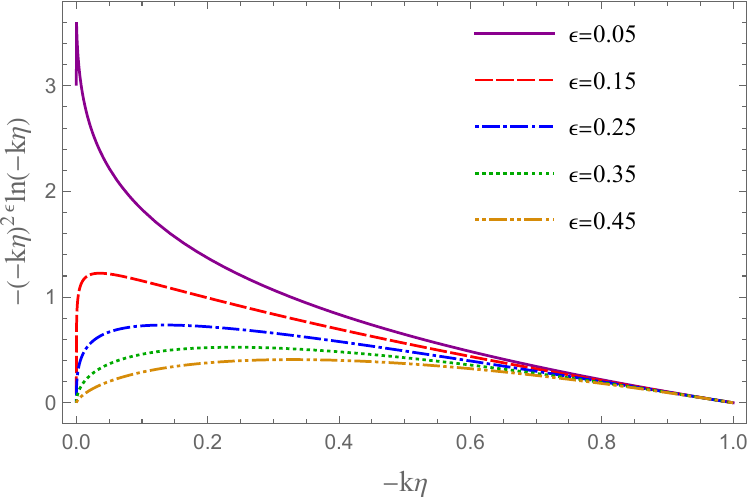}
	\caption{The curve of $\delta n_B$  as a function of $-k\eta$.}
	\label{f1}
\end{figure}
Figure \ref{f1} shows the curve of $\delta n_B$  as a function of $-k\eta$, with different values of $\epsilon$ represented. It can be seen that $\delta n_B$ first increases and then decreases, indicating that the deviation of large-scale magnetic fields from a scale-invariant spectrum first grows and then diminishes.

At a fixed moment in time, the qualitative behavior of the deviation in the electric field spectrum index is the same as in the case of $\gamma = 1$. In contrast, for the magnetic field spectrum index deviation, the behavior of $n_B$ follows the same pattern as the curve with $\epsilon = 0.25$ in Figure \ref{f1} — that is, there exists a characteristic scale $k_{\rm max}$ at which the deviation of the magnetic field spectrum from a scale-invariant spectrum reaches its maximum. Numerical results show that this scale is:
\begin{equation}
	\label{e108}
	-k_{\rm max}\eta\sim e^{-2}\Rightarrow k^{-1}_{\rm max}\sim e^2|\eta|
\end{equation}
indicating that this characteristic scale gradually decreases as inflation progresses.

\subsection{Comparison with Planck 2018 Results}
Exponential potential inflation does not agree well with observational data. Therefore, in this section, we discuss several inflationary models that are more consistent with observations.
We focus on the case where the magnetic field has a scale-invariant spectrum ($\gamma = 2$). In this case, we have:
\begin{equation}
	\label{ea8}
	\delta n_B\Big|_{\rm inf}\sim \frac{3H^2}{\sqrt{2}\pi^2V_{\phi}}\Big|_{\rm inf}\ln(-k\eta_{\rm inf})\sqrt{k}
\end{equation}
The slow-roll parameter is defined as:
\begin{equation}
	\label{ea9}
	\epsilon=\frac{m_{pl}^2}{16\pi}\left(\frac{V_{\phi}}{V}\right)^2
\end{equation}
Under the slow-roll approximation, 

\begin{equation}
	\label{ea10}
	V\sim\frac{3-\epsilon}{8\pi}m_{pl}^2H^2
\end{equation}
then,
\begin{equation}
	\label{ea11}
	V_{\phi}\sim\pm\frac{m_{pl}}{2\sqrt{\pi}}\sqrt{\epsilon}(3-\epsilon)H^2
\end{equation}
The "$\pm$" sign depends on the sign of $V_\phi$. Consider a pivot scale $k_*$, and assume that inflation starts when the $k_*$ mode exits the horizon. The number of e-folds from horizon exit to the end of inflation is denoted by $N$, then $\ln(-k\eta_{\rm inf})\sim N$. Insert this and \eqref{ea11} into \eqref{ea8} we have:
\begin{equation}
	\label{ea12}
	\delta n_B\Big|_{\rm inf}\sim \frac{3\sqrt{2k_*}}{\pi^{3/2}m_{pl}}\varsigma
\end{equation}
where
\begin{equation}
	\label{ea13}
	\varsigma:=\mp\frac{N}{\sqrt{\epsilon}(3-\epsilon)}
\end{equation}

The quantity $\varsigma$ is related to the slow-roll parameter $\epsilon$ and the number of e-folds $N$. It characterizes the deviation of the primordial magnetic field spectrum from scale invariance.
	
For different inflationary models, the relation between the slow-roll parameter $\epsilon$ and the number of e-folds $N$ varies. Both $\epsilon$ and $N$ can be connected to observable quantities such as the scalar spectral index $n_S$ and the tensor-to-scalar ratio $r$, where $n_S$ characterizes the spectral tilt of scalar perturbations, and $r$ represents the ratio of tensor to scalar perturbation amplitudes. In this way, the quantity $\varsigma$ can be related to $n_S$ and $r$ within different inflationary scenarios.
\begin{figure}[htbp]
	\includegraphics[width=\columnwidth]{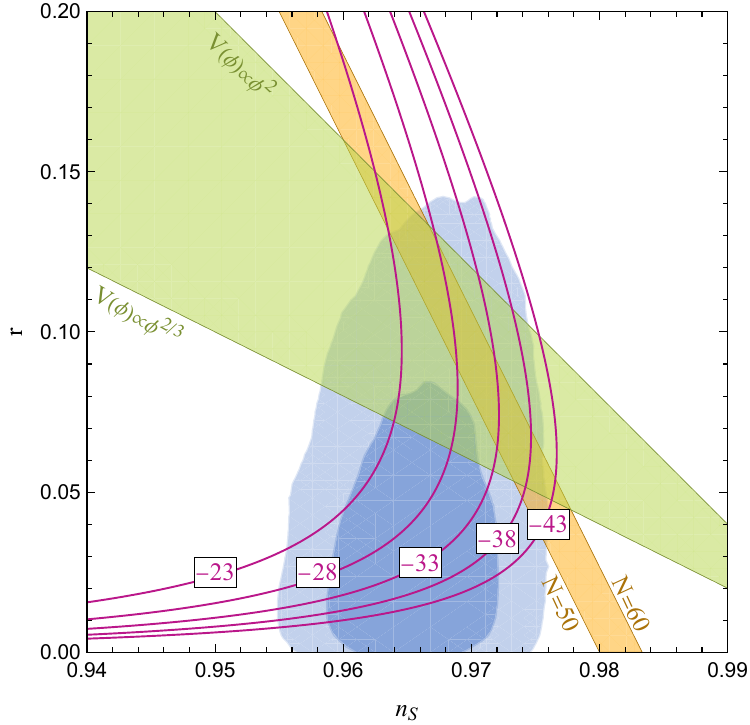}
	\caption{Contour lines of $\varsigma$ on the $n_S$–$r$ plane for the power-law potential $V(\phi)\propto \phi^p$.
		}
	\label{fa8}
\end{figure}

In Fig.\ref{fa8}, contour lines of $\varsigma$ on the $n_S$–$r$ plane for the power-law potential $V(\phi)\propto \phi^p$ are shown. The blue region denotes the observational constraints from Planck 2018. The yellow region corresponds to the theoretical range with $N \in [50,60]$, the green region represents the theoretical range with $p \in [2/3, 2]$.
In the intersection of these two regions, the power-law potential model shows good agreement with the Planck 2018 observational results \cite{2020_collaboration}. 
The purple solid lines indicate the contours of $\varsigma$. 
From the figure, the range of $\varsigma$ in the intersection regions is approximately between $-43$ and $-23$. The negative values arise because the power-law potential model belongs to the class of large-field inflation models.

\begin{figure}[htbp]
	\includegraphics[width=\columnwidth]{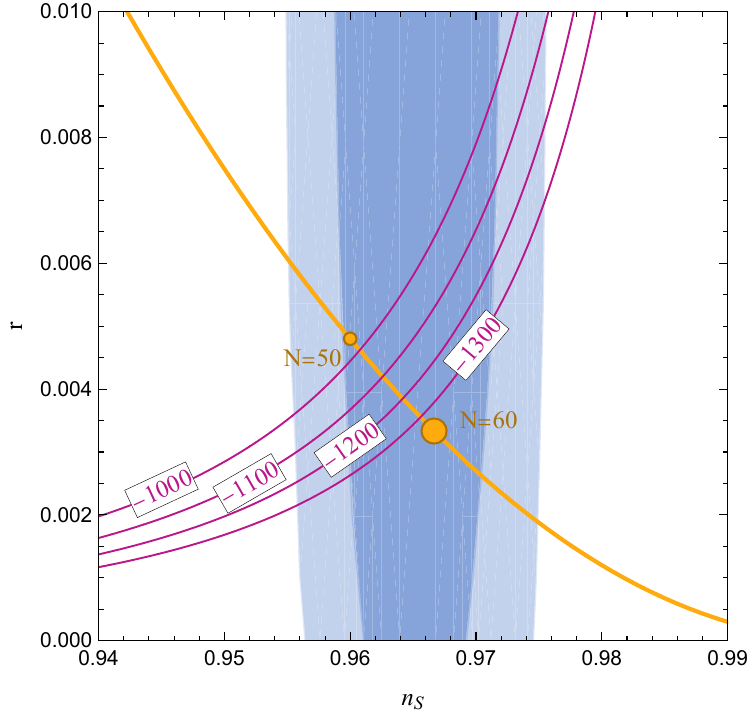}
	\caption{Contour lines of $\varsigma$ on the $n_S$–$r$ plane for the Starobinsky potential $V(\phi)\propto \phi^p$.}
	\label{fa9}
\end{figure}

Fig.\ref{fa9} shows the contour lines of $\varsigma$ for the Starobinsky potential, where the yellow curves represent the theoretically predicted region. The number of e-folds $N$ can be considered as a parameter along the curves. From the figure, it can be seen that within the range $N \in [50, 60]$, $\varsigma$ takes values approximately between $-1300$ and $-1000$. It is noteworthy that although both are large-field models, the values of $\varsigma$ under the Starobinsky potential differ significantly from those under the power-law potential. This implies that the value of $\varsigma$ can be used to distinguish different inflationary models.

\begin{figure}[htbp]
	\includegraphics[width=\columnwidth]{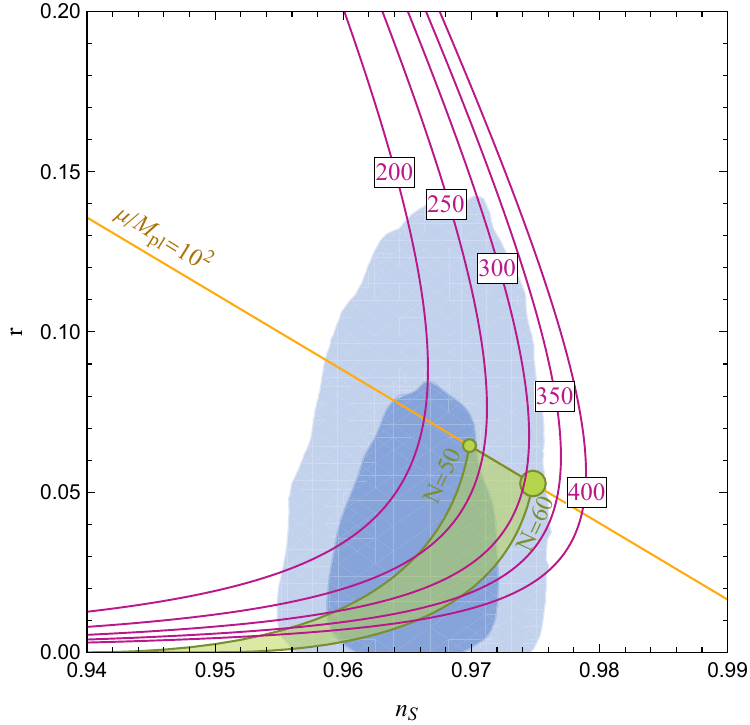}
	\caption{Contour lines of $\varsigma$ on the $n_S$–$r$ plane for the hilltop potential $V(\phi)\propto 1-(\phi/\mu)^4$.}
	\label{fa10}
\end{figure}

Fig.\ref{fa10} shows the contour lines of $\varsigma$ for the hilltop potential $V(\phi)\propto1-(\phi/\mu)^p$. The hilltop model contains two parameters: $p$ and $\mu$. For simplicity, we consider only the quartic hilltop case with $p = 4$. The green region in the figure represents the theoretical parameter space with $N \in [50, 60]$ and $\mu / M_{pl} \in [10^{-2}, 10^{2}]$. In this region, the model agrees well with the Planck 2018 observational data. As shown in the figure, $\varsigma$ takes values approximately in the range of 250 to 400. The positive sign of $\varsigma$ reflects the fact that the hilltop model belongs to the class of small-field inflation models.

\section{Summary\label{s5}}
In this paper, we have investigated the influence of inhomogeneous perturbations on the PMF.
 We first generalize the Ratra model to a general spacetime background and analyze the constraint algebra structure of the electromagnetic field. In a spacetime background with inhomogeneous spatial perturbations, the standard Coulomb gauge does not automatically satisfy the Gauss constraint. Therefore, to construct magnetogenesis models during inflation with inhomogeneous perturbations, the Coulomb gauge must be generalized. Unlike the conventional treatment of the electromagnetic field, in order to conveniently solve the constraint conditions together with Maxwell’s equations, we do not solve directly for the vector potential of the magnetic field. Instead, we first solve for its conjugate momentum.

When determining the initial conditions for the mode function $\varpi$ of the conjugate momentum, we find that the coupling function introduces corrections to the initial conditions of $\varpi$. These corrections are negligible at the early stages of inflation but grow over time and become dominant in the late stages of inflation. This directly leads to a difference between the electromagnetic power spectrum obtained in this model and that of conventional inflationary magnetogenesis models. Assuming a power-law form for the coupling function, the conventional model yields a scale-invariant electric field spectrum when $\gamma = -3, 2$, and a scale-invariant magnetic field spectrum when $\gamma = -2, 3$. However, after taking into account the corrections to the initial conditions of $\varpi$ caused by the coupling function, the electromagnetic field achieves a scale-invariant spectrum at $\gamma = -2, 1$ for the electric field, and at $\gamma = 2$ for the magnetic field.

We have also analyzed the backreaction problem. As in conventional inflationary magnetogenesis models, avoiding backreaction imposes an upper bound on the energy density of the inflaton field. We find that this upper bound lies above the energy scale required by Big Bang nucleosynthesis, indicating that inflationary magnetogenesis models without backreaction problems are indeed possible.

Another challenge in calculating the influence of inhomogeneous perturbations on the primordial magnetic field spectrum lies in computing the convolution between the vector potential and the inhomogeneous perturbations. To address this, we divide the convolution into two parts and apply large-scale and small-scale approximations separately in each part. Using this method, we derive the impact of inhomogeneous perturbations of the inflaton on the large-scale electromagnetic power spectrum. To quantify this impact, we introduce $\Delta_E$ and $\Delta_B$, and we found that their ratio under the slow-roll approximation is independed of inflationary model. We also found that when the electric field has a scale-invariant spectrum, the influence of inhomogeneous perturbations on the electric and magnetic power spectrum is proportional, with the proportionality factor depending only on the scale under consideration. When the magnetic field has a scale-invariant spectrum, the influence of inhomogeneous perturbations on the magnetic power spectrum increases relative to that of the electric field as inflation proceeds.

We also have presented the behavior of $\Delta_E$ and $\Delta_B$ for different values of $\gamma$ and $\epsilon$ under the exponential potential. We find that $\Delta_E$ and $\Delta_B$ exhibit oscillations before horizon exit, with their amplitudes increasing as $\gamma$ and $\epsilon$ become larger.

From the perspective of the spectrum index, we find that the effect of inhomogeneous perturbations depends on the sign of $V_\phi$. For the electric field spectrum, regardless of whether it is scale-invariant, $V_\phi > 0$ shifts the spectrum index toward the blue end, with the degree of shift decreasing with scale and becoming smaller as inflation progresses, and vice versa. For the magnetic field, the behavior differs significantly between scale-invariant and non-scale-invariant cases: for non-scale-invariant spectra, $ V_\phi > 0 $ also shifts the spectral index toward the blue end, but the degree of shift increases with scale and becomes smaller as inflation progresses.

For scale-invariant magnetic field spectrum, the impact of inhomogeneous perturbations on the spectrum index is more complex. Due to the presence of logarithmic factors, the spectral index shifts toward the red end when $ V_\phi > 0 $.

We have presented the constraints imposed by the Planck 2018 observational results on the deviation of the primordial magnetic field spectral index from exact scale invariance in different inflationary models. We find that the degree of deviation differs significantly among these models. This indicates that observations of nearly scale-invariant PMF spectrum can be used to distinguish between large-field and small-field inflation models. Moreover, there exists a time-dependent scale at which the deviation from scale invariance of the magnetic field is maximized. As inflation proceeds, this scale decreases, indicating that the effect of inhomogeneous perturbations on scale-invariant magnetic field spectrum shifts toward smaller scales.

\section*{Acknowledgments}
This work was supported by the Fundamental Research Funds for the Central Universities of
Ministry of Education of China under Grants No. 3132018242,
the Natural Science Foundation of Liaoning Province of China under Grant No.20170520161 and the National Natural Science Foundation of China under Grant No.11447198 (Fund of theoretical physics).



\bibliography{references}

\end{document}